\documentclass[11pt]{article}
\usepackage{amsfonts,amsmath,amsthm,array,amssymb}
\usepackage{graphicx}
\usepackage{graphics}
\usepackage{multicol}
\usepackage{cite}
\usepackage{tikz}
\usepackage{relsize}
\usepackage[normalem]{ulem}
\usepackage{color}
\usepackage{epstopdf}

\begin{document}


\title {Substance Abuse via Legally Prescribed Drugs: The Case of Vicodin in the United States}
\author{Wendy K. Caldwell$^{1}$, Benjamin Freedman$^{2}$, Luke Settles$^{3}$, Michael M. Thomas$^{4}$,\\ Anarina Murillo$^{5}$, Erika Camacho$^{5,6}$, Stephen Wirkus$^{5,6}$}
\date{}
\maketitle
\begin{center}
\footnotesize $^{1}$ Department of Mathematics, University of Tennessee-Knoxville, Knoxville, TN\\
\vspace{.05in}
\footnotesize $^{2}$ Department of Mathematics, Bucknell University, Lewisburg, PA\\
\vspace{.05in}
\footnotesize $^{3}$ Department of Mathematics and Statistics, Southern Illinois University Edwardsville, Edwardsville, IL\\
\vspace{.05in}
\footnotesize $^{4}$ Department of Mathematics and Statistics, Kennesaw State University, Kennesaw, GA\\
\vspace{.05in}
\footnotesize $^{5}$ Mathematical, Computational, and Modeling Sciences Center, Arizona State University, Tempe, AZ\\
\vspace{.05in}
\footnotesize $^6$ School of Mathematical and Natural Sciences, Arizona State University at West Campus, Glendale, AZ
\end{center}
\vspace{.25in}
\hrule
\vspace{.25in}


\begin{abstract}
{Vicodin is the most commonly prescribed pain reliever in the United States. Research indicates that there are two million people who are currently abusing Vicodin, and the majority of those who abuse Vicodin were initially exposed to it via prescription. Our goal is to determine the most effective strategies for reducing the overall population of Vicodin abusers. More specifically, we focus on whether prevention methods aimed at educating doctors and patients on the potential for drug abuse or treatment methods implemented after a person abuses Vicodin will have a greater overall impact. We consider one linear and two non-linear compartmental models in which medical users of Vicodin can transition into the abuser compartment or leave the population by no longer taking the drug. Once Vicodin abusers, people can transition into a treatment compartment, with the possibility of leaving the population through successful completion of treatment or of relapsing and re-entering the abusive compartment. The linear model assumes no social interaction, while both non-linear models consider interaction. One considers interaction with abusers affecting the relapse rate, while the other assumes both this and an additional interaction between the number of abusers and the number of new prescriptions. Sensitivity analyses are conducted varying the rates of success of these intervention methods measured by the parameters to determine which strategy has the greatest impact on controlling the population of Vicodin abusers. These results give insight into the most effective method of reducing the number of people who abuse Vicodin. From these models, we determine that manipulating parameters tied to prevention measures has a greater impact on reducing the population of abusers than manipulating parameters associated with treatment. We also note that increasing the rate at which abusers seek treatment affects the population of abusers more than the success rate of treatment itself.}
\end{abstract}


\section{Introduction}
Among medically accessible pain relievers, Vicodin is the most widely prescribed in the United States \cite{mechcatie2009advisory}. Although it comprises 4\% of the world's population, the United States uses 99\% of the world's supply of hydrocodone, the narcotic agent in Vicodin \cite{manchikanti2007national}. Vicodin is a Schedule III narcotic and is a combination of hydrocodone, an opioid analgesic, and acetaminophen, the active ingredient in Tylenol. The increase in prescriptions that has taken place in the last two decades has resulted in a corresponding growth in Vicodin abuse \cite{mccabe2007motives}. Abuse rates increased from 7\% in 1993 to 16\% in 2003 \cite{comer2008growth}. Research indicates that the most common path to becoming a Vicodin abuser begins with a prescription. Most abusers obtain the drug via prescription, whether it be their own or a prescription of a friend or relative \cite{micahael2013how}. Vicodin abuse can cause a number of dangerous side effects: liver failure, difficulty breathing, jaundice, slowed heart rate, seizures, and  death \cite{prescrip2013effects}.  Many prescribers of Vicodin are unaware of its potential for chemical and physical dependence \cite{oneill2013}. Most abusers who initially start using Vicodin because of a doctor's prescription are not informed of the risks for drug dependence \cite{compton2006major}. Although treatment methods and prevention programs exist, they have varying success rates and ignore the shifting needs of the drug-abusing population as it ages. Research indicates that treatment should be tailored to different demographics rather than a one-size-fits-all approach \cite{cicero2012patterns}. \\
\indent In 1999, approximately 9 million Americans admitted to using prescription drugs for non-medical reasons \cite{manchikanti2007national}. Consequently, population-level models for prescription drug abuse exist \cite{sung2005nonmedical, maxwell2006trends, compton2006major}. Because Vicodin relapse dynamics are comparable to those in smoking tobacco as well as some diseases, we are using the framework of an epidemiological model, considering the level and means of Vicodin use for our compartments \cite{sleeper2006}. However, unlike many epidemiological models, we do not consider social interaction necessary to entering or leaving the compartments. Because most Vicodin abusers are introduced to the drug via prescription and not through experimentation with other users, there is no assumed interaction between the medical user and abuser compartments. Those who come to Vicodin through other means most frequently obtain the drug from someone with a prescription, so the population of non-prescribed recreational users can be lessened by decreasing the number of prescribed abusers \cite{mccabe2007motives}.\\
\indent Educating medical professionals has proven successful in limiting the number of Vicodin prescriptions \cite{oneill2013}. However, 40\% of medical professionals indicated they had received no training on the risks of Vicodin dependency. Abuse prevention measures focused on increasing education of physicians, pharmacists, and the public have been generally neglected \cite{manchikanti2007national}. Education of physicians regarding Vicodin abuse prevention has been neglected, and doctors and patients are often left without necessary information to manage the risk of taking the drug. This information increases the ability to recognize abuse and can aid in preventing its spread \cite{census}. A study of pharmacists in the United States and Canada indicated that nearly 90\% of pharmacists had refused to fill a prescription for a patient when there were concerns of drug abuse, and more than 75\% had attempted to contact the prescribing physician when they had such concerns \cite{azzopardicontrolling}. A program implemented in California to educate prescribers of Vicodin on the risks associated with the drug led to a 95\% decrease in the number of Vicodin prescriptions \cite{oneill2013}.\\
\indent In the following sections, we develop three mechanistic mathematical models for a population introduced to Vicodin by prescription and their dynamics and transition through the stages of medical use, drug abuse, and treatment. We model relapse into the abuse compartment in two ways: two non-linear models incorporating social interaction and one linear Compartmental Vicodin Transition (CVT) Model without it. In Section \ref{sec:linmod}, we include the analysis of the CVT Model. In Section 3, which contains the non-linear models, we consider the rate of entrance into the population as a constant through the Social Interaction with Constant Prescription Rate (SIC) Model and also varying according to the population of Vicodin abusers through the Social Interaction with Abuse-Dependent Prescription Rate (SIAD) Model. By analyzing three models addressing these characteristics, we are able to better identify the main drivers of abuse and relapse and how to better prevent their occurrences. We analyze the impact of our parameters and initial conditions on the solutions and the impact of parameters on our steady-state system to determine whether focusing on abuse prevention methods is treatment methods would be more effective in reducing the number of Vicodin abusers in this population.


\section{Compartmental Vicodin Transition Model}\label{sec:linmod}
\begin{figure}
\begin{center}
\includegraphics[height=3.75cm,width=12cm]{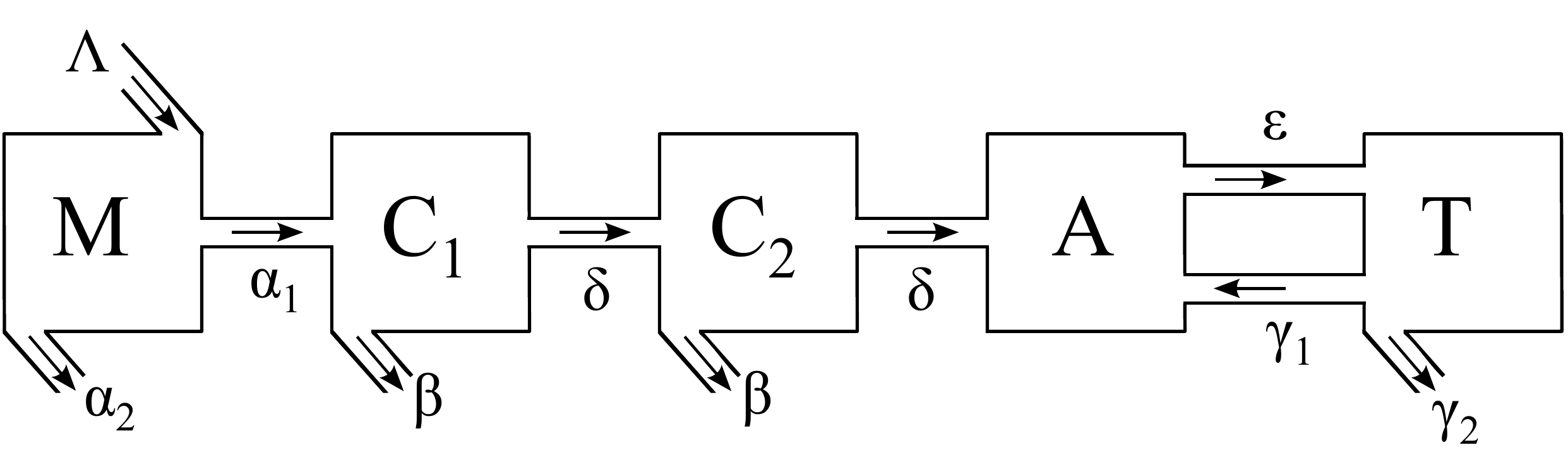} 
\caption {\small CVT Model. \smaller{This figure shows the linear representation of a population of Vicodin patients as they transition through chronic use, abuse, treatment, and possible relapse.}}
\label{Linear Compartmental Model}
\end{center}

\end{figure}

\indent In this model, we consider a population of individuals initially prescribed Vicodin by a medical professional and classify them according to level of Vicodin use. The first compartment, consisting of acute medical users $(M)$, is the one into which people immediately enter when prescribed the narcotic. If the supply of Vicodin is only for up to three months, patients leave the population. If not, they enter a chronic compartment ($C_1$). They transition to the $C_2$ compartment if they continue to take Vicodin for medical reasons. If individuals in $C_2$ begins taking the drug either recreationally or in a manner inconsistent with the prescribed dosage, they become members of the abuse compartment $(A)$. Otherwise, if the patients stop taking Vicodin, they leave $C_2$ and exit the population. Once in the abusive compartment, individualsl can either remain there or seek treatment \cite{volkow2005prescription, herzanek2009don}. Individuals in the treatment compartment $(T)$ can either leave the population through successful treatment or re-enter the $A$ compartment through relapse. The model considers a 40-month time period. Figure ~\ref{Linear Compartmental Model} shows the flow diagram for this model, along with associated linear rates between compartments. The parameters are defined in Table ~\ref{Parameter Definitions}, and the compartments are described by equations (\ref{dMdtlin}) - (\ref{dTdtlin}). An in-depth explanation of parameter values can be found in Appendix IV.  For the purposes of this model, we assume 30 days represents one month. \\
\indent We are able to obtain a lower bound of 0.00125\%, derived from the number of people who used prescription opioids for non-medical use, and an upper bound of 0.126\%, derived from the number of people who sought treatment for abuse \cite{CDCPolicyImpact}. We thus conclude that the number of abusers who die from overdosing on Vicodin is not statistically significant and can be neglected for this model.  For an explanation of all parameters excluded from this model, see Appendix IV.

\begin{table}[htbp]
\caption{\small Parameter Explanations}
\label{Parameter Definitions}
\vspace{.1in}
\hspace{-.4in}
\begin{tabular}{|l|l|l|l|}\hline
Parameter & Definition & Unit & Value\\ \hline \hline
$\Lambda$ & rate of new medical Vicodin users entering the population & $\frac{\mbox{people}}{\mbox{month}}$ & [2671212, 3303044] \\ \hline
$\alpha_1$ & rate of acute users becoming chronic users & $\frac{1}{\mbox{month}}$ & [0.175, 0.240] \\ \hline
$\alpha_2$ & rate of acute users ending Vicodin treatment & $\frac{1}{\mbox{month}}$ & $1.762\alpha_1 \leq\alpha_2\leq7.850\alpha_1$ \\ \hline
$\beta$ & rate of chronic users ending Vicodin treatment & $\frac{1}{\mbox{month}}$ & $0.205\beta \leq \delta \leq 0.513\beta$\\ \cline{1-3}
$\delta$ & rate of chronic users moving to next compartment & $\frac{1}{\mbox{month}}$ & $0.0862-\beta \leq \delta \leq 0.256-\beta$ \\ \hline
$\epsilon$ & rate of abusers entering treatment for Vicodin abuse & $\frac{1}{\mbox{month}}$ & [.014,.042]\\ \hline
$\gamma_1^*$ & relapse rate (CVT Model) & $\frac{1}{\mbox{month}}$ & [.046,.45]\\ \hline
$\gamma_2$ & successful treatment rate & $\frac{1}{\mbox{month}}$ & [.038,.55]\\ \hline
\end{tabular}
\smaller *For both the SIC and SIAD Models, the units of $\gamma_1$ change to $\frac{1}{\mbox{people}\times \mbox{month}}$, where people is defined by the population of the United States in recent years, and its value range is [$1.26\times10^{-10}$, $1.50\times10^{-9}$]. Refer to Appendix IV for derivations and references.
\end{table}


\begin{eqnarray}
\frac{dM}{dt}&=&\Lambda-(\alpha_1+\alpha_2)M \label{dMdtlin}\\
\frac{dC_1}{dt}&=&\alpha_1M-(\delta+\beta)C_1 \label{dC1dtlin}\\
\frac{dC_2}{dt}&=&\delta C_1-(\delta+\beta)C_2 \label{dC2dtlin}\\
\frac{dA}{dt}&=&\delta C_2+\gamma_1T-\epsilon A \label{dAdtlin}\\
\frac{dT}{dt}&=&\epsilon A-(\gamma_1+\gamma_2) T\label{dTdtlin}
\end{eqnarray}
\indent Equation (\ref{dMdtlin}) calculates the population of the acute medical user compartment ($M$) by taking the inflow of new acute medical users per month ($\Lambda$) and subtracting the population that exits $M$ per month (the $M$ population multiplied by the sum of both exit rates, $\alpha_1$, and $\alpha_2$). Equation (\ref{dC1dtlin}) represents the population of the first chronic user compartment ($C_1$). This is calculated by taking the flow into $C_1$, $\alpha_1M$, and subtracting the outflow, ($\delta$ + $\beta$)$C_1$. Equation (\ref{dC2dtlin}) represents the population of $C_2$ and is calculated by taking the flow into $C_2$, $\delta$$C_1$ and subtracting the flow out of $C_2$, ($\delta$ + $\beta$)$C_2$. Equation (\ref{dAdtlin}) represents the population of the $A$ compartment. This is calculated by taking $\delta$$C_2$ and $\gamma_1$$T$, the entrances into $A$, and subtracting $\epsilon$$A$, the exit from $A$. Finally, Equation (\ref{dTdtlin}) represents the $T$ population. We calculate this by taking the entrance into $T$, $\epsilon$$A$, and subtracting the exits from $T$, ($\gamma_1$ + $\gamma_2$)$T$.

In effort to understand the long-term dynamics of this system, we compute the equilibrium point of the system, denoted $(M^*, C_1^*, C_2^*, A^*, T^*)$ and given by:
\begin{eqnarray}
M^*&=&\frac{\Lambda}{\alpha_1+\alpha_2}\nonumber \\
C_1^*&=&M^*\left(\frac{\alpha_1}{\delta+\beta}\right)\nonumber \\
C_2^*&=&C_1^*\left(\frac{\delta}{\delta+\beta}\right)\nonumber \\
A^*&=&T^*\left(\frac{\gamma_1+\gamma_2}{\epsilon}\right)\nonumber \\
T^*&=&C_2^*\left(\frac{\delta}{\gamma_2}\right)\nonumber
\end{eqnarray}
which is globally stable, meaning that regardless of parameter values and initial populations, the end result over time is the equilibrium point (see Appendix I). The equilibrium point $M^*$ corresponds to the carrying capacity of the acute medical user population. That is, the values of the acute user compartment will reach the steady state ($M^*$) independent of the initial population size. Note, the equilibrium of each of the other four classes is dependent on the carrying capacity, and the steady state of each class is defined recursively in terms of the steady states of other classes by multiplying by the entrance rate and the average waiting time in each compartment (i.e. $\frac{\alpha_1}{\delta+\beta}$). This demonstrates the significance of the steady state in propagating through the long-run behavior of the model, as long run behavior is determined by the long run behavior of other classes, and all depend on the carrying capacity. $M^*$ is the ratio of new Vicodin patients to the rates at which people leave the M compartment. Note that this ratio influences the magnitudes of all other steady states. Therefore, $M^*$ is important to the overall size of the population. 


\subsection{Simulations of CVT Model}
\begin{figure} [http]
\begin{centering}
\includegraphics[scale=0.7]{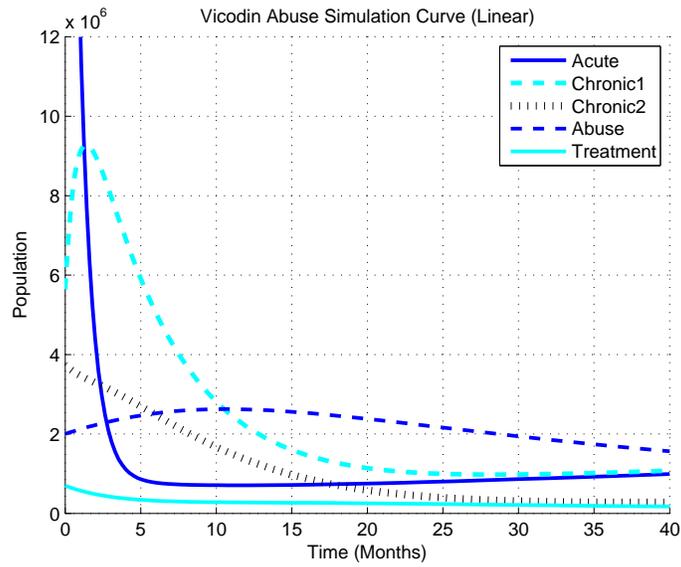}
\caption{\small CVT Model Simulation with Arbitrary Parameter Values. \smaller{The plot above displays a 40-month period with arbitrary parameters selected from the acceptable ranges according to data. It predicts that although the first 12 months show a peak in the population of Vicodin abusers, the number of abusers decreases in the next 24 months.}}
\label{VicodinSimulationCurve1}
\end{centering}
\end{figure}

\begin{figure} [http]
\begin{center}
\includegraphics[scale=0.7]{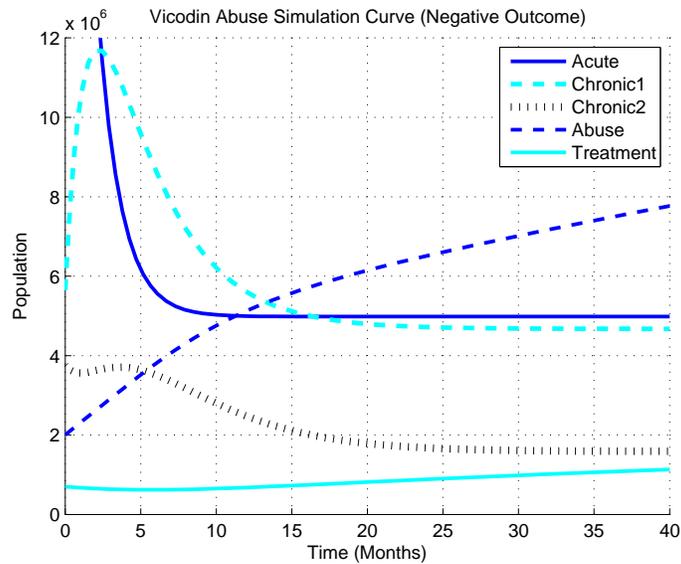}
\end{center}
\caption{\small CVT Model Pessimistic Curve. \smaller{This plot displays the scenario in which all of the parameters assume values of the least desirable outcomes. In this case, the abuse population grows by a large degree, so it is necessary to enhance Vicodin abuse prevention and treatment methods.}}
\label{pessimist}
\end{figure}

\begin{figure} 
\begin{centering}
\includegraphics[scale=0.7]{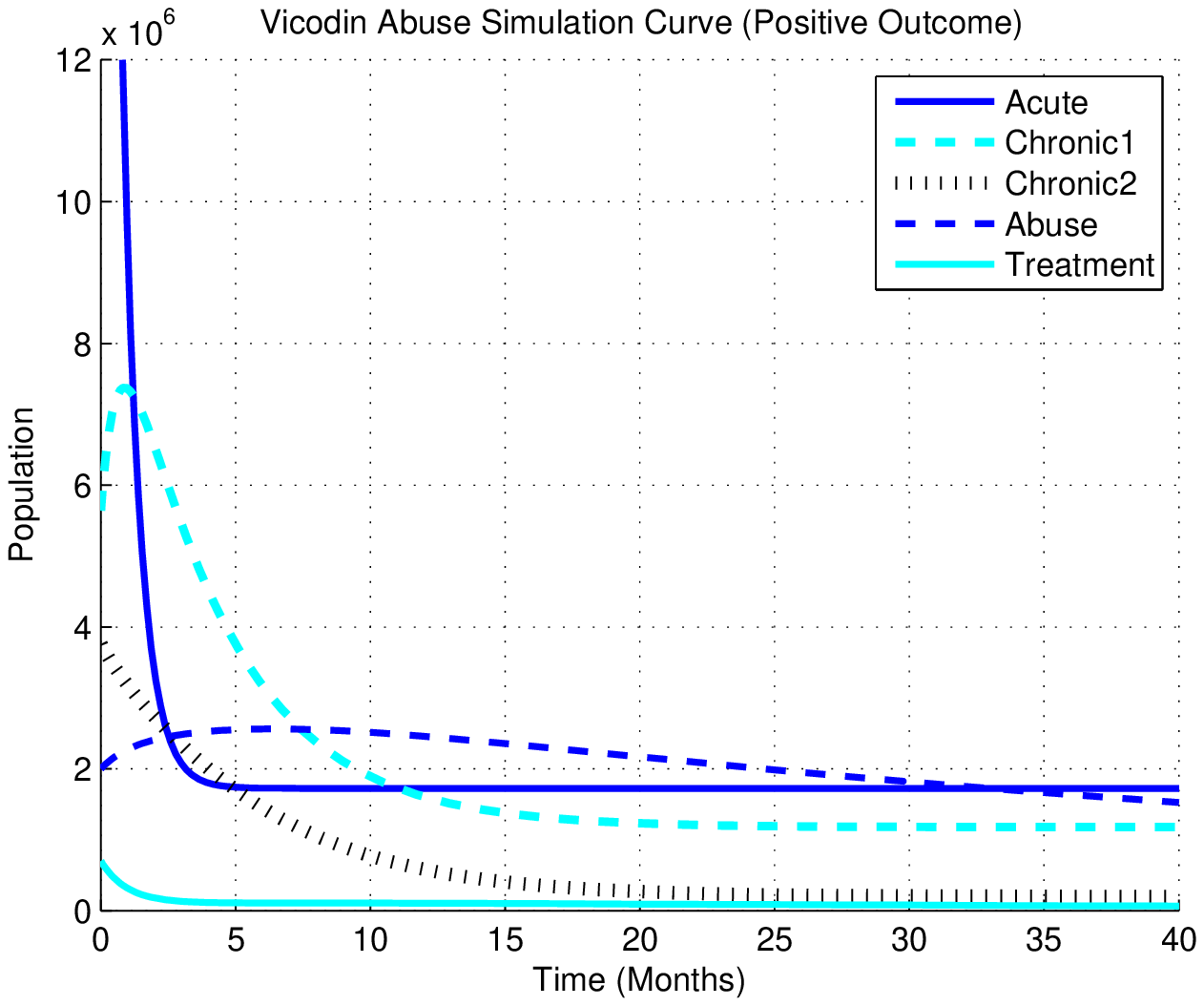}
\caption{\small{CVT Model Optimistic Curve}. \smaller{The figure shows the behaviors of the populations of the compartments if all of the parameters assume their most desirable values within our range. The population of users is kept to a minimum.}}
\label{optimist}
\end{centering}
\end{figure}

Figures~\ref{VicodinSimulationCurve1}-\ref{linsens} were created with initial conditions obtained from the Substance Abuse and Mental Health Services Administration, Michael's House, and \textit{TIME} \cite{samhsa2012results, szalavitz2013fda, VicodinRehabMH2013}. Initially, there are 37.6 million people in the acute compartment, 5.64 million in the first chronic compartment ($C_1$), 3.76 million in the second chronic compartment ($C_2$), 2 million in the abuser compartment ($A$), and 700,000 in the treatment compartment ($T$) \cite{VicodinRehabMH2013, szalavitz2013fda, samhsa2012results}. The values of the parameters are assumed to remain constant over time because we do not consider exogenous perturbation that would change the parameters over time (see Table~\ref{Parameter Definitions} for parameter definitions). Using the function ODE15 in Matlab 2013a, we numerically estimate the number of people in our compartments over a course of 40 months. This time frame is realistic for practical applications as it allows enough time for the system to stabilize and provides a reasonable timeline for policy implementations and further study.

\begin{table}[htbp]
\caption{\small Simulation Parameter Values}
\label{Parameter Values (Linear)}
\begin{center}
\begin{tabular}{|l|l|l|l|l|l|l|l|l|}\hline
&$\Lambda$ & $\alpha_1$ & $\alpha_2$ & $\beta$ & $\delta$ & $\epsilon$ &  $\gamma_1$ &$\gamma_2$ \\ \hline
Arbitrary&$3000000$ & $.220$ & $.950$ & $.140$ & $.0500$ & $.0300$ &  $.240$ &$.293$ \\ \hline
Pessimist&$3303044$ & $.240$ & $.423$ & $.169$ & $.0869$ & $.014$ &  $.038$ &$.0458$ \\ \hline
Optimist&$2671212$ & $.175$ & $1.374$ & $.213$ & $.0436$ & $.042$ &  $.45$ &$.55$ \\ \hline
\end{tabular}
\end{center}
\smaller For parameter explanations, see Table \ref{Parameter Definitions}.
\end{table}

\begin{table}[htbp]
\caption{\small Simulation Initial Conditions}
\label{Initial Condition}
\begin{center}
\begin{tabular}{|l|l|l|l|l|}\hline
$M_{0}$ & $C_{1_0}$ & $C_{2_0}$ & $A_{0}$ & $T_{0}$ \\ \hline
$37,600,000$ & $5,640,000$ & $3,760,000$ & $2,000,000$ & $700,000$  \\ \hline
\end{tabular}
\end{center}
\smaller{These are the initial coniditions used for all simulations. For parameter explanations, see Table \ref{Parameter Definitions}}.
\end{table}

\indent The simulation curves in Figure~\ref{VicodinSimulationCurve1} show a steep decrease in the number of acute users. This is due to the exit rate of $M$ being significantly greater than the exit rates of the other compartments, which is consistent with the data \cite{sullivan2008trends}.  It displays the populations of the respective compartments over the course of 40 months. The population of compartment $C_1$ experiences a spike to 9 million people in the initial two months, but then decreases to 1 million in the following five months. After 40 months, there are approximately 1.7 million abusers, a decrease of 300,000 from the inital population. The model predicts that the Vicodin abuser population increases and exceeds that of the other compartments after 10 months for the selected parameters. This is a fairly positive outcome, because the sum of the three medical populations is greater than the abuse population. 

Figure~\ref{pessimist} assumes the worst case for parameter values, in which rates of transition into abuse and into relapse are at their maximums. Thus, compared with the previous graph, Figure \ref{pessimist} has a larger abuser population. This is noteworthy as it indicates the possibility of a large, steadily increasing abuser population. At the end of the time interval, there are a predicted 8 million individuals in the abuser compartment. Figures~\ref{VicodinSimulationCurve1} and~\ref{optimist} have early small peaks in abuse followed by slow decrease. Figure \ref{pessimist} shows no peak or decrease in the populaion of $A$, indicating a steady rise in the population of abusers. There still fails to be a time in which the abuser population exceeds the sum of the medical populations. This plot shows that there may be a need for a change in Vicodin abuse prevention and/or treatment policy.

Figure~\ref{optimist} makes opposing assumptions to those of Figure~\ref{pessimist}. The parameters for movement into the substance-abusing population ($\alpha_{1}$, $\delta$ , $\epsilon$, $\gamma_{1}$) are at their lower bounds, and parameters for exiting the total population ($\beta$, $\alpha_{2}$, $\gamma_{2}$) are high. The figure shows that the populations are generally lower, and the abuse compartment decreases in population size. This is indicative of a declining $A$ population that eventually drops below the $M$ population. This decrease is the result of an elevated rate of exit from the abuser compartment. After approximately 10 months, the number of abusers begins to decline, and after 30 months, it falls below the number of acute users ($M$). This plot shows that our estimated parameter range includes a case in which abuse is declining, so prevention or treatment may already be having a limited controlling effect.\\
\indent The graph with arbitrary parameters predicts a decrease in Vicodin abuse over the next $40$ months after increasing to a peak in $10$ months. The most desirable set of parameters predicts a similar result but sooner, reaching the peak around six months. The least desirable set of parameters predicts an increase in abuse over $40$ months.


\subsection{Sensitivity Analysis of the CVT Model}\label{subsec:senslinmod}

\indent Sensitivity analysis involves a numerical method of solving the adjoint equations to analyze the influence of all parameters in the model in addition to examining the normalized sensitivity indices of the equilibrium points of the system. Refer to Appendix VI for the derivation of the adjoint equations. Specifically, we focus on how a small perturbation of each parameter influences the population of the abuser compartment over time. The sensitivity of the parameters of the system is illustrated in Figure~\ref{linsens}. The normalized sensitivity of selected parameters of the equilibrium is discussed in the next section.\\
\indent The results from the sensitivity analyses enable us to measure the degree to which each of the rates affects the number of abusers. From this information, we know which rates should be changed in order to decrease the abuser population as much as possible. From this, we can determine the most effective method (prevention, treatment, etc.) of controlling the abusers.


\subsection{Analysis of Normalized Sensitivity Indices of the CVT Model}\label{subsec:normsenslin}
Recall that $A$ is the compartment of abusers, and we are interested in how treatment or prevention affects this compartment. We consider reducing $\delta$, the rate at which chronic patients become Vicodin abusers, as prevention and reducing $\gamma_1$, the relapse rate, as indicative of successful treatment. The parameter $\delta$, the transition rate between $C_2$ and $A$, is also the transition rate between $C_1$ and $C_2$. For simplicity, we assume these two rates to be equal and give them the same corresponding parameter ($\delta$). This is an assumption that we make because we are concerned with the behavior of $A$, and the division of the chronic classes is based on the length of time spent in the chronic class. Additionally, $\beta$ and $\gamma_2$ are indicators of prevention and treatment, respectively. We analyze the sensitivity of the equilibrium solution of abusers, $A^*$:
\begin{eqnarray}
A^*&=&\left(\frac{\Lambda}{\alpha_1+\alpha_2}\right)\left(\frac{\alpha_1}{\delta+\beta}\right)\left(\frac{\delta}{\delta+\beta}\right)\left(\frac{\delta}{\gamma_2}\right)\left(\frac{\gamma_1+\gamma_2}{\epsilon}\right). \nonumber
\end{eqnarray}
\indent In finding the sensitivity of $A^*$ with respect to each parameter, we compute the partial derivative of $A^*$ with respect to the parameter and divide that by the ratio of $A^*$ to the parameter.  This results in the value of the ratio, $\frac{\mbox{\% change in $A^*$}}{\mbox{\% change in the parameter}}$, or the elasticity of $A^*$ with respect to each parameter. So for each percent change in a certain parameter, the percent change in $A^*$ corresponds to the change in the parameter multiplied by this ratio of percent changes. We select arbitrary values of percent change that we fluctuate parameters by in order to observe the resulting percent change in $A^*$. These changes are standard through analysis of each parameter so we can observe similarities and differences of each $A^*$ with regards to the same change in various parameters.\\
\indent Analyzing the percent change of $A^*$ with respect to the percent change in $\gamma_1$:
\begin{eqnarray}
\frac{\partial A^*}{\partial\gamma_1}\frac{\gamma_1}{A^*} = \frac{\gamma_1}{\gamma_1+\gamma_2}. \nonumber
\end{eqnarray}

\indent Analyzing the percent change of $A^*$ with respect to the percent change in $\gamma_2$:
\begin{eqnarray}
\frac{\partial A^*}{\partial\gamma_2}\frac{\gamma_2}{A^*} = \frac{-\gamma_1}{\gamma_1+\gamma_2} \nonumber
\end{eqnarray}

\indent Examining the percent change of $A^*$ with respect to the percent change in $\delta$:
\begin{eqnarray}
\frac{\partial A^*}{\partial\delta}\frac{\delta}{A^*} = \frac{2\beta}{\delta+\beta} \nonumber
\end{eqnarray}

\indent Examining the percent change of $A^*$ with respect to the percent change in $\beta$:
\begin{eqnarray}
\frac{\partial A^*}{\partial\beta}\frac{\beta}{A^*} = \frac{-2\beta}{\delta+\beta} \nonumber
\end{eqnarray}

\indent Analysis  of $\epsilon$, which is the rate at which abusers go to treatment, can determine if increasing $\epsilon$ affects the $A$ compartment. Thus, examining the percent change of $A^*$ with respect to the percent change in $\epsilon$:
\begin{eqnarray}
\frac{\partial A^*}{\partial\epsilon}\frac{\epsilon}{A^*} = -1 \nonumber
\end{eqnarray}

\indent Analyzing the rate of percent change of $A^*$ with respect to the percent change in $\Lambda$, the rate of new Vicodin-prescribed patients per month:
\begin{eqnarray}
\frac{\partial A^*}{\partial\Lambda}\frac{\Lambda}{A^*} = 1 \nonumber
\end{eqnarray}
\indent We first choose arbitrary values of our parameters from the estimated ranges in Table~\ref{Parameter Definitions} ($\gamma_1=0.200$, $\gamma_2=0.250$, $\delta=0.053$, $\beta=0.150$). The results are shown in Figure \ref{linNormsens}. Additional results can be found in Appendix V.
\begin{figure} [h]
\begin{center}
\includegraphics[scale=1]{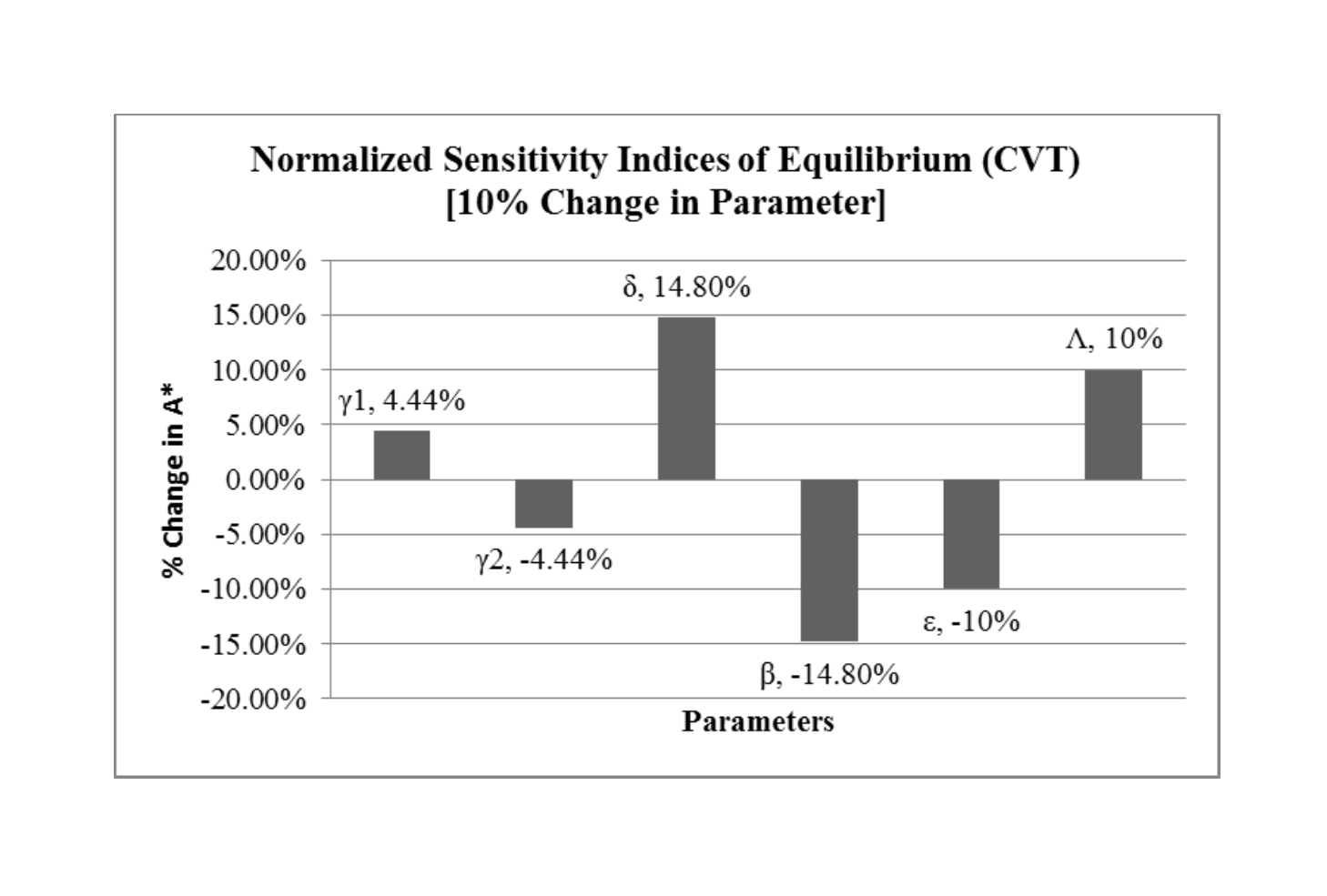}
\end{center}
\vspace{-.5in}
\caption{\small Normalized Sensitivity Indices of CVT Model. \smaller{ The magnitude of these percentages measures the effect of the indicated parameter on $A^*$. For tables of values of pertubations other than 10\%, see Appendix V.}}
\label{linNormsens}
\end{figure}
From Figure~\ref{linNormsens}, we conclude that prevention ($\delta$ and $\beta$) affects $A^*$ more than the other parameters. The treatment parameters ($\gamma_1$ and $\gamma_2$) lead to a smaller percentage change in $A^*$. A percent change in $\delta$ and $\beta$ yields an even larger percent change in number of abusers. For example, a $10\%$ change in $\delta$ yields a $14.8\%$ change in $A^*$ while a $10\%$ change in $\gamma_1$ yields a $4.44\%$ change in $A^*$. Note that $\delta$ and $\beta$ have the same magnitude of effect, as do $\gamma_1$ and $\gamma_2$.


\subsection{Adjoint Sensitivity Analysis of the CVT Model}\label{subsec:adjointlin}
We conduct sensitivity analysis of our parameters to determine those that have the greatest effect on the population of the $A$ compartment. We utilize an adjoint method for sensitivity analysis to find the sensitivity equations needed, which yields the same result as forward sensitivity analysis (see Appendix III). Recall the system of equations of the CVT Model is:
\begin{eqnarray}
\frac{dM}{dt}&=&\Lambda-(\alpha_1+\alpha_2)M \nonumber \\
\frac{dC_1}{dt}&=&\alpha_1M-(\delta+\beta)C_1 \nonumber \\
\frac{dC_2}{dt}&=&\delta C_1-(\delta+\beta)C_2 \nonumber \\
\frac{dA}{dt}&=&\delta C_2+\gamma_1T-\epsilon A \nonumber \\
\frac{dT}{dt}&=&\epsilon A-(\gamma_1+\gamma_2) T. \nonumber 
\end{eqnarray}
\indent Let $\dot{M}$ denote $\frac{dM}{dt}$, $\dot{C_1}$ denote $\frac{dC_1}{dt}$, $\dot{C_2}$ denote $\frac{dC_2}{dt}$, $\dot{A}$ denote $\frac{dA}{dt}$, and $\dot{T}$ denote $\frac{dT}{dt}$. Using the adjoint sensitivity method \cite{BradleyPDE}, we rewrite the system of ordinary differential equations: 
\begin{center}
$\vec{F}(t,$ $\vec{x}$, $\dot\vec{x}$, $\vec{p}) = \begin{bmatrix}
	\dot{M}-\Lambda+(\alpha_1+\alpha_2)M \\
	\dot{C_1}-\alpha_1M+(\delta+\beta)C_1 \\
	\dot{C_2}-\delta C_1+(\delta+\beta)C_2 \\
	\dot{A}-\delta C_2-\gamma_1T+\epsilon A \\
	\dot{T}-\epsilon A+(\gamma_1+\gamma_2) T
\end{bmatrix} = 0_{5\times1}$,
\end{center}
\indent where
\vspace{0.1in}
$\vec{x}^T = \begin{bmatrix}
	M & C_1 & C_2 & A & T
\end{bmatrix}$ is the vector containing the different compartment populations over time, and 
$\vec{x}(0)^T = \begin{bmatrix}
	M(0) & C_1(0) & C_2(0) & A(0) & T(0)
\end{bmatrix}$ is the vector containing the initial population sizes of each compartment. \\
\indent We define our parameter vector $\vec{p}$ such that \setcounter{MaxMatrixCols}{14}
$\vec{p}^T = \begin{bmatrix}
	\Lambda & \alpha_1 & \alpha_2 & \delta & \beta & \epsilon & \gamma_1 & \gamma_2 & u_1 &  u_2 &  u_3 &  u_4 &  u_5
\end{bmatrix}$, where $u_i$ for $i$=1,...,5 represents our initial condition parameters for which the sensitivity indices will be computed. For the purpose of calculating the sensitivity of our solutions to our initial conditions, we define
\begin{center}
$\vec{y}(0) = \begin{bmatrix}
	M(0)(1-u_1) \\
	C_1(0)(1-u_2) \\
	C_2(0)(1-u_3) \\
	A(0)(1-u_4) \\
	T(0)(1-u_5)
\end{bmatrix}.$ 
\end{center} 
$\vec{y}$(0) is the vector that contains the $u_i$ (for $i$=1,...,5) percent change of the initial population sizes of the compartments. \\
\indent Because we want to minimize the population of abusers ($A$), we define the objective function $A(\vec{x}, \vec{p}) = \int^T_0 g(\vec{x}, t, \vec{p}) dt = \int^T_0 \dot{A} dt$. 
We want to analyze this, because we are interested in how much the abuser population ($A$) is affected by small changes in the parameters and the initial population size of each compartment ($\vec{p}$). \\
\indent Following the second step of the algorithm for computing the sensitivity equations (see Appendix VI) \cite{BradleyPDE}, the adjoint is $g_x + \lambda^T(F_x-\dot{F}_{\dot{x}})-\dot{\lambda}^T F_{\dot{x}} = 0$, where: 
\begin{itemize}
\item $\lambda$ is the Lagrange multiplier, and $\lambda^T = \begin{bmatrix}
		\lambda_1 & \lambda_2 & \lambda_3 & \lambda_4 & \lambda_5
	\end{bmatrix}$
	
\item $g_x$ is the partial of $A$ with respect to the population sizes of the compartments ($\vec{x}$):
\begin{center}
$g_x = \frac{\partial\dot{A}}{\partial\vec{x}} = \begin{bmatrix}
	\frac{\partial\dot{A}}{\partial M} & \frac{\partial\dot{A}}{\partial C_1} & \frac{\partial\dot{A}}{\partial C_2} & \frac{\partial\dot{A}}{\partial A} & \frac{\partial\dot{A}}{\partial T}
\end{bmatrix} = \begin{bmatrix}
	0 & 0 & \delta & -\epsilon & \gamma_1 
\end{bmatrix}$
\end{center}

\item $F_x$ is the partial of our system of ODEs ($F$) with respect to $\vec{x}$:
\begin{center}
$F_x = \begin{bmatrix}
	(\alpha_1+\alpha_2) & 0 & 0 & 0 & 0 \\
	-\alpha_1 & (\delta+\beta) & 0 & 0 & 0 \\
	0 & -\delta & (\delta+\beta) & 0 & 0 \\
	0 & 0 & -\delta & \epsilon & -\gamma_1 \\
	0 & 0 & 0 & -\epsilon & (\gamma_1+\gamma_2)
\end{bmatrix}$
\end{center}

\item $\dot{\vec{x}}$ is the derivative of compartment vector ($\vec{x}$) with respect to time $\left(\dot{\vec{x}}^T = \begin{bmatrix}
	\dot{M} & \dot{C_1} & \dot{C_2} & \dot{A} & \dot{T}
\end{bmatrix}\right)$.

\item $F_{\dot{x}}$ is the partial of our system of ODEs ($F$) with respect to the $\dot{\vec{x}}$:
\begin{center}
$F_{\dot{x}} = \begin{bmatrix}
	1 & 0 & 0 & 0 & 0 \\
	0 & 1 & 0 & 0 & 0 \\
	0 & 0 & 1 & 0 & 0 \\
	0 & 0 & 0 & 1 & 0 \\
	0 & 0 & 0 & 0 & 1
\end{bmatrix}$
\end{center}

\item $\dot{F}_{\dot{x}}$ is the parital of the derivative of our ODE system ($\dot{F}$) with respect to $\dot{\vec{x}}$ and $\dot{F}_{\dot{x}} = 0_{5\times5}$.
\end{itemize}

\indent Now, solving for the $\lambda^T(F_x-\dot{F}_{\dot{x}})$ part of the adjoint, we note that $F_x-\dot{F}_{\dot{x}} = F_x$. Therefore, multliplying the transpose of the Lagrange multiplier vector by the partial of our system of ODEs with respect to the compartmental populations yields: \\
$\lambda^T F_x = \begin{bmatrix}
	(\alpha_1+\alpha_2)\lambda_1-\alpha_1\lambda_2 & (\delta+\beta)\lambda_2-\delta\lambda_3 & (\delta+\beta)\lambda_3-\delta\lambda_4 & \epsilon\lambda_4-\epsilon\lambda_5 & \gamma_1\lambda_4+(\gamma_1+\gamma_2)\lambda_5
\end{bmatrix}$. \\
\indent For the final section of the adjoint ($\dot{\lambda}^T F_{\dot{x}}$), the product of the transpose of the Lagrange multiplier vector and the partial of our system of ODEs with respect to the derivative of the compartmental populations is:
$\dot{\lambda}^T F_{\dot{x}} = \begin{bmatrix}
	\dot{\lambda_1} & \dot{\lambda_2} & \dot{\lambda_3} & \dot{\lambda_4} & \dot{\lambda_5}
\end{bmatrix}$. \\
\indent Combining the results from above, we obtain the adjoint equation:
\begin{eqnarray}
(\alpha_1+\alpha_2)\lambda_1-\alpha_1\lambda_2-\dot{\lambda}_1&=&0 \nonumber \\
(\delta+\beta)\lambda_2-\delta\lambda_3-\dot{\lambda}_2&=&0 \nonumber \\
\delta+(\delta+\beta)\lambda_3-\delta\lambda_4-\dot{\lambda}_3&=&0 \nonumber \\
-\epsilon+\epsilon\lambda_4-\epsilon\lambda_5-\dot{\lambda}_4&=&0 \nonumber\\
\gamma_1-\gamma_1\lambda_4+(\gamma_1+\gamma_2)\lambda_5-\dot{\lambda}_5&=&0 \nonumber
\end{eqnarray}
with initial conditions $\lambda_i$(T)=0, for $i$ = 1, ...,5. \\
\indent Now, we simultaneously solve the initial value problem: $F=0$, \\$\vec{x}(0)^T = \begin{bmatrix} M(0) & C_1(0) & C_2(0) & A(0) & T(0) \end{bmatrix}$. Also, we solve the general sensitivity equation  
$\frac{dA}{dp}=\int_0^T (g_p+\lambda^TF_p) dt+\lambda^T F_{\dot{x}}|_{t=0} \vec{y}_{x(0)}^{~-1}\vec{y_p}$, where:
\newcommand\bigzero{\makebox(0,0){\text{\Huge0}}}
\begin{itemize}
\item $g_p$ is the partial of $\dot{A}$ with respect the the vector of parameters ($\vec{p}$):
\begin{center}
 $g_p = \begin{bmatrix}
	0 & 0 & 0 & C_2 & 0 & -A & T & 0 & 0 & 0 & 0 & 0 & 0
\end{bmatrix}$
\end{center}

\item $F_p$ is the partial of our system of ODEs ($F$) with respect to the the parameters ($\vec{p}$):
\begin{center}
$F_p=\begin{bmatrix}
-1&M&M&0&0&0&0&0& & & & &\\
0&-M&0&C_1&C_1&0&0&0& & & & & \\ 
0&0&0&C_2-C_1&C_2&0&0&0 & &\bigzero& & &\\ 
0&0&0&-C_2&0&A&-T & 0 & & & 5\times5 & &\\ 
0&0&0&0&0&-A&T&T & & & & &
\end{bmatrix}$.
\end{center}

\item $\vec{y}_{p}$ is the partial of the percent change in initial compartment populations ($\vec{y}$) with respect to the parameters ($\vec{p}$):
\begin{center}
$\vec{y}_{p}=\begin{bmatrix} &&&&&&&-M(0)&0&0&0&0\\&&&&&&&0&-C_1(0)&0&0&0\\&&&\bigzero&&&&0&0&-C_2&0&0\\&&&&5\times8&&&0&0&0&-A(0)&0\\&&&&&&&0&0&0&0&-T(0) \end{bmatrix}.$
\end{center}

\item $\vec{y}_{x(0)}^{~-1}$ is the inverse of the partial of the percent change in initial compartment populations ($\vec{y}$) with respect to the initial population vector ($\vec{x}$(0)): \\
$\vec{y}_{x(0)}=\begin{bmatrix} 1-u_1&0&0&0&0 \\ 0&1-u_2&0&0&0 \\ 0&0&1-u_3&0&0 \\ 0&0&0&1-u_4&0 \\ 0&0&0&0&1-u_5 \end{bmatrix}$,
$\vec{y}_{x(0)}^{~-1}=\begin{bmatrix} \frac{1}{1-u_1}&0&0&0&0 \\ 0&\frac{1}{1-u_2}&0&0&0 \\ 0&0&\frac{1}{1-u_3}&0&0 \\ 0&0&0&\frac{1}{1-u_4}&0 \\ 0&0&0&0&\frac{1}{1-u_5} \end{bmatrix}$.

\item All previously seen expressions ($\lambda^T$, $F_{\dot{x}}$) have the same definitions from above.
\end{itemize}

\indent Within the integrand of the general sensitivity equation, we have the product of the Lagrange multiplier vector ($\lambda^T$) and the partial derivative of our system of ODEs with respect to the the parameters ($F_p$), which is:
\begin{center}
$\lambda^T F_p = \begin{bmatrix} v & w \end{bmatrix}$, where
\begin{eqnarray}
v &=& \begin{bmatrix}-\lambda_1 & (\lambda_1-\lambda_2) M & \lambda_1M & (\lambda_2-\lambda_3)C_1+(\lambda_3-\lambda_4)C_2\end{bmatrix}\nonumber \\
w &=& \begin{bmatrix}\lambda_2C_1+\lambda_3C_2 & (\lambda_4-\lambda_5)A & (-\lambda_4+\lambda_5)T \end{bmatrix}\nonumber
\end{eqnarray}
\end{center}
 
\indent Multiplying the the Lagrange multiplier vector ($\lambda^T$) by the partial of our system of ODEs with respect to the derivative of the compartment vector with respect to time ($F_{\dot{x}}$) with the inverse of the partial of the percent change in initial compartment populations with respect to the initial population vector ($\vec{y}_{x(0)}^{~-1}$), and with the partial of the percent change in initial compartment populations with respect to the parameters ($\vec{y}_{p}$), we have: 
\begin{center}
$F_{\dot{x}}|_{t=0} \vec{y}_{x(0)}^{~-1}\vec{y_p}=
\begin{bmatrix} &&&&&&&&\frac{-M(0)}{1-u_1}&0&0&0&0\\&&&&&&&&0&\frac{-C_1(0)}{1-u_2}&0&0&0 \\ &&&\bigzero&&&&&0&0&\frac{-C_2(0)}{1-u_3}&0&0 \\ &&&&5\times8&&&&0&0&0&\frac{-A(0)}{1-u_4}&0 \\ &&&&&&&&0&0&0&0&\frac{-T(0)}{1-u_5} \end{bmatrix}$,
\end{center}
$\lambda^TF_{\dot{x}}|_{t=0} \vec{y}_{x(0)}^{~-1}\vec{y_p} = \begin{bmatrix}
	0 & 0 & 0 & 0 & 0 & 0 & 0 & 0 & -\lambda_1\frac{M(0)}{1-u_1} & -\lambda_2\frac{C_1(0)}{1-u_2} & -\lambda_3\frac{C_2(0)}{1-u_3} & -\lambda_4\frac{A(0)}{1-u_4} & -\lambda_5\frac{T(0)}{1-u_5}\end{bmatrix}$. \\
\indent Thus, from $\frac{dA}{dp}$, we add $g_p$ and $\lambda^T F_p$ together within the integrand and then add $\lambda^TF_{\dot{x}}|_{t=0} \vec{y}_{x(0)}^{~-1}\vec{y_p}$ outside of it. Now, we have the sensitivity equations:
\begin{eqnarray}
\frac{\partial A}{\partial \Lambda} &=& \int_0^T -\lambda_1dt \nonumber\\
\frac{\partial A}{\partial \alpha_1} &=& \int_0^T (\lambda_1-\lambda_2) M dt \nonumber\\
\frac{\partial A}{\partial \alpha_2} &=&\int_0^T \lambda_1Mdt \nonumber\\
\frac{\partial A}{\partial \delta} &=& \int_0^T(\lambda_2-\lambda_3)C_1+(\lambda_3-\lambda_4+1)C_2dt \nonumber\\
\frac{\partial A}{\partial \beta} &=& \int_0^T \lambda_2C_1+\lambda_3C_2 dt\nonumber\\
\frac{\partial A}{\partial \epsilon} &=& \int_0^T (\lambda_4-\lambda_5-1)A dt \nonumber\\
\frac{\partial A}{\partial \gamma_1} &=& \int_0^T (-\lambda_4+\lambda_5+1)T dt \nonumber\\
\frac{\partial A}{\partial \gamma_2} &=& \int_0^T \lambda_5Tdt \nonumber\\
\frac{\partial A}{\partial u_1} &=& -\lambda_1\frac{M(0)}{1-u_1} \nonumber\\
\frac{\partial A}{\partial u_2} &=& -\lambda_2\frac{C_1(0)}{1-u_2} \nonumber\\
\frac{\partial A}{\partial u_3} &=& -\lambda_3\frac{C_2(0)}{1-u_3} \nonumber\\
\frac{\partial A}{\partial u_4} &=& -\lambda_4\frac{A(0)}{1-u_4} \nonumber\\
\frac{\partial A}{\partial u_5} &=& -\lambda_5\frac{T(0)}{1-u_5}. \nonumber
\end{eqnarray}
\indent By varying the upper limit of integration, T, from [0, 60], we are able to get the sensitivity of the population of the abusers ($A$) with respect to each parameter over the first 60 months using MATLAB. See Figure~\ref{linsens}.


\subsection{Results and Conclusions of the CVT Model}\label{subsec:resultslin}
\begin{figure} [htbp]
\begin{center}
\includegraphics[scale=.8]{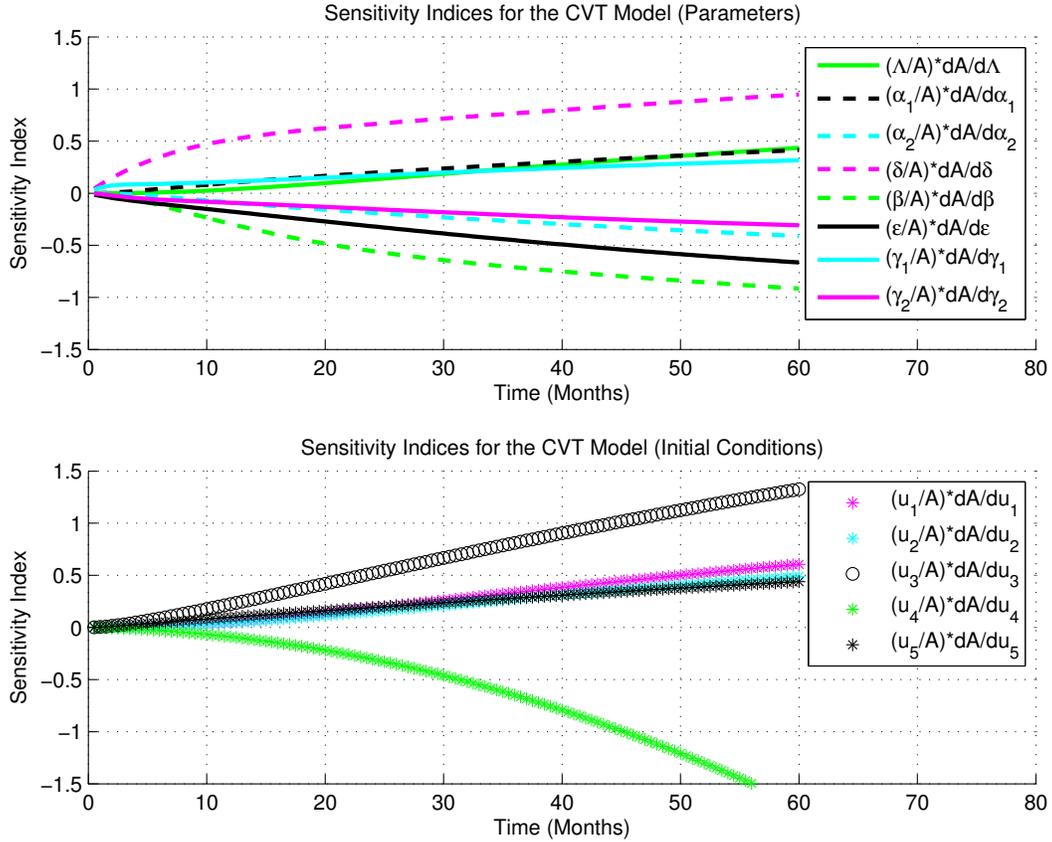}
\end{center}
\caption{\small Sensitivity of the CVT Model. \smaller{These figures illustrate the sensitivity indices ($\frac{\Lambda}{A} \frac{\partial A}{\partial \Lambda}$, etc.) and indicate the degree to which the parameters are correlated to the number of people in $A$. The plots show that improvements in prevention (decreasing $\delta$ and/or increasing $\beta$) and a higher intervention rate (increasing $\epsilon$) have the greatest decreasing effect on the number of people in the abuse compartment. In the bottom graph, the most influential initial population is the abuser compartment ($u_4$). (Note: in the top graph, $\frac{\alpha_1}{A} \frac{\partial A}{\partial \alpha_1}$ and $\frac{\Lambda}{A} \frac{\partial A}{\partial \Lambda}$ overlap after approximately 35 months; in the bottom graph, $u_1$, $u_2$, and $u_5$ overlap; for initial conditions refer to Table \ref{Initial Condition})}}
\label{linsens}
\end{figure}
\begin{figure} [htbp]
\begin{center}
\includegraphics[scale=.8]{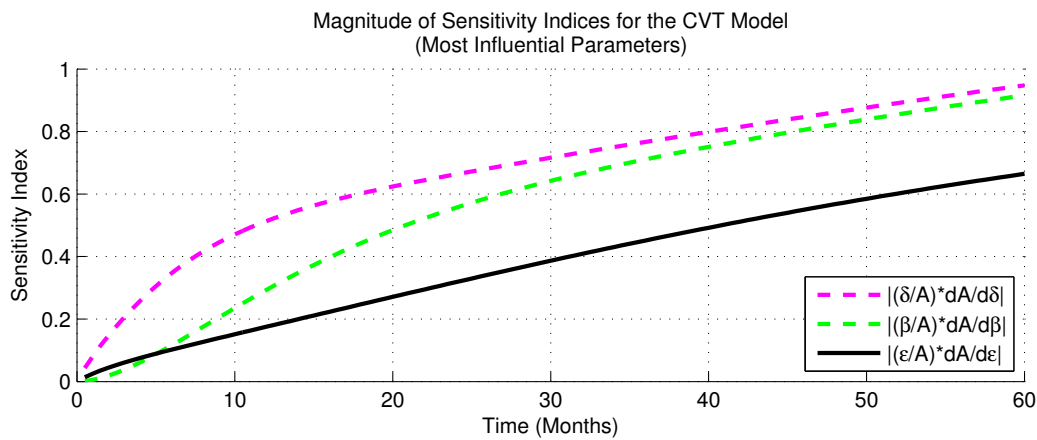}
\end{center}
\caption{\small Sensitivity Index Magnitudes Comparison for the CVT Model. \smaller{This plot compares the magnitudes of the sensitivity indices of the most influential parameters for the linear model from Figure \ref{linsens}. Starting with the strongest influence on the size of the abuser compartment, we have the rate at which chronic users become abusers ($\delta$), the rate at which chronic users stop taking Vicodin ($\beta$), and the rate at which abusers enter treatment ($\epsilon$). The sensitivity indices of the initial conditions were not considered, because the current population values cannot be changed. The indices stabilize after approximately 200 months.}}
\label{linsensmag}
\end{figure}

\indent The sensitivity analysis allows us to observe the long-term behavior of the system, assuming no interaction between compartments. We determine how influential each parameter is in affecting the population of $A$ over time and whether or not each parameter is positively or negatively correlated with $A$. Figure~\ref{linsens} and \ref{linsensmag} show that increasing abuse-prevention efforts (i.e., decreasing $\delta$ and/or increasing $\beta$) is the strongest method for decreasing the number of people in the Vicodin abuser compartment. At 60 months, the sensitivity indices are 0.9478 and -0.9160 for $\delta$ and $\beta$, respectively. The intervention rate ($\epsilon$) is the next most influential in reducing the population of $A$ with a sensitivity index of -0.6646. The initial value of $A$ has a strong inverse relationship with the number of abusers over time, as demonstrated in the lower portion of Figure~\ref{linsens}. All other initial conditions have a direct relationship, with $C_2$ having the greatest influence among those. However, the current populations of each compartment, which we take as our initial conidtions, are not changeable. Introducing a non-linear model will enable us to observe social interaction, specifically those in the $A$ and $T$ compartments, to determine if this impacts the population of $A$.\\
\indent The parameters and initial conditions of the simulation curves of the CVT Model were derived from data. However, there are some limitations to the curves themselves. For example, there is a sharp decrease in the number of acute medical users ($M$) during the first few months of the simulation. This is not the data suggests. In Figure \ref{pessimist}, the population of the abuse compartment outnumbers the other compartment populations. Recall that the total number of medical users is the sum of $M$, $C_1$, and $C_2$. However, abusers do not actually outnumber medical users.

\section{Non-linear Models}\label{nonlinmod}
In order to address the fact that people in treatment are more prone to relapse in the event that they come into contact with those who are still abusing Vicodin while in treatment, we introduce two models with abuser-treatment interaction terms, a more realistic interpretation of the data.


\subsection{Social Interaction with Constant Prescription Rate (SIC) Model}
This  model incorporates social interaction between the abusers ($A$) and those in treatment ($T$). We assume that those in the $T$ compartment who interact with those in the $A$ compartment are more likely to re-enter the $A$ compartment than those who do not interact, which is consistent with the data. A new range for the parameter $\gamma_1$ results from dividing $\gamma_1$ from the CVT Model by the population of the United States (300 million people). This new range is $\gamma_1 \in [1.26 \times 10^{-10}, 1.50 \times 10^{-9}]$. This model is based on studies that show social interaction between abusers and those in treatment hinders recovery and increases the chance for relapse \cite{abusegroup2009treatment, caronstats2013}.

\begin{figure}[htbp]
\begin{center}
\includegraphics[height=3.25cm,width=12cm]{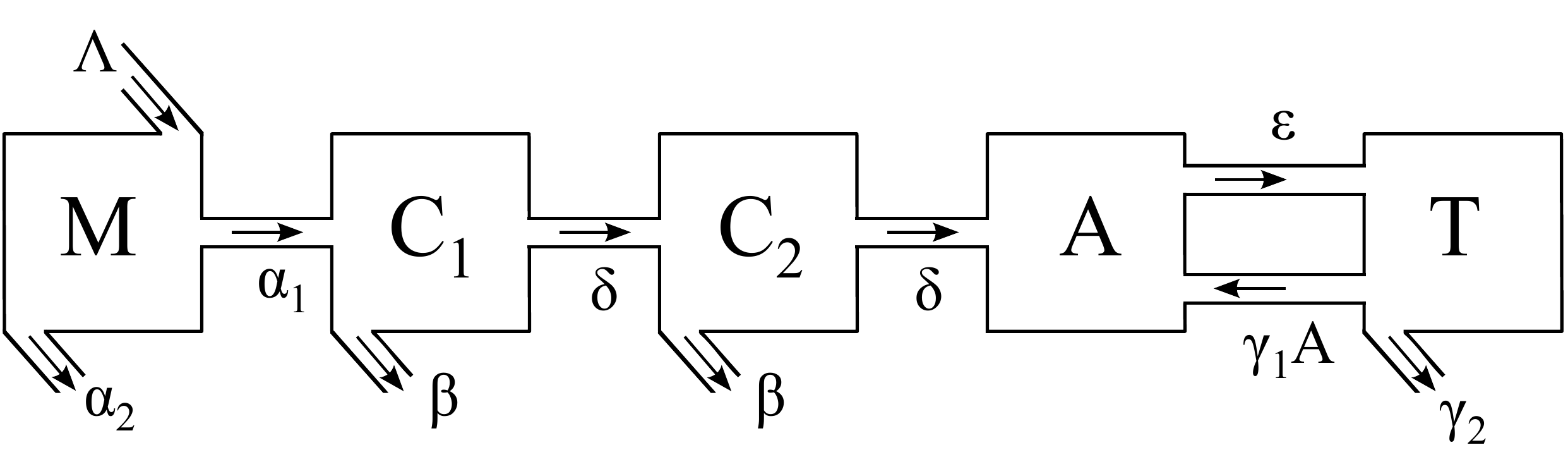} 
\caption{\small Social Interaction with Constant-Prescription Rate (SIC) Model. \smaller{This figure shows a model with social interaction between abusers ($A$) and those in treatment ($T$).}}
\label{Social Interaction Model}
\end{center}
\end{figure}

The governing dynamics of this system are given by: 
\begin{eqnarray}
\frac{dM}{dt}&=&\Lambda-(\alpha_1+\alpha_2)M\label{dMdtsocial} \\
\frac{dC_1}{dt}&=&\alpha_1M-(\delta+\beta)C_1\label{dC1dtsocial} \\
\frac{dC_2}{dt}&=&\delta C_1-(\delta+\beta)C_2\label{dC2dtsocial} \\
\frac{dA}{dt}&=&\delta C_2+\gamma_1AT-\epsilon A\label{dAdtsocial} \\
\frac{dT}{dt}&=&\epsilon A-\gamma_1AT-\gamma_2T\label{dTdtsocial}.
\end{eqnarray}
\indent This model is identical to the CVT Model with the exception of the interaction-influenced relapse rate. This term increases the accuracy of our model \cite{caronstats2013, marlatt2005relapse}.

The equilibrium point, denoted $(M^*, C_1^*, C_2^*, A^*, T^*)$, is:
\begin{eqnarray}
M^*&=&\frac{\Lambda}{\alpha_1+\alpha_2}\nonumber \\
C_1^*&=&M^*\left(\frac{\alpha_1}{\delta+\beta}\right)\nonumber \\
C_2^*&=&C_1^*\left(\frac{\delta}{\delta+\beta}\right)\nonumber \\
A^*&=&\frac{\gamma_2T^*}{\epsilon-\gamma_1T^*} \nonumber \\
T^*&=&C_2^*\left(\frac{\delta}{\gamma_2}\right).\nonumber
\end{eqnarray}
Observe that $A^*$ is positive only when $T^* < \frac{\epsilon}{\gamma_1}$. We are only interested in cases where $A^*$ is positive because that is when it is biologically relevant.\\
\indent Let $\gamma_L$ be the rate of relapse from the CVT Model, and note that $\gamma_1=\frac{\gamma_L}{N}$, where $N$ is the population of the United States. This condition becomes $\gamma_L\frac{T^*}{N} < \epsilon$. If the product of the relapse rate from the CVT model and the proportion of Americans in treatment is less than the rate at which abusers seek treatment, $A^*$ is biologically relevant.

To assess stability of the equilibrium point, we linearize the system by looking at the Jacobian, the matrix of partial derivatives. If all eigenvalues of the Jacobian evaluated at the equilibrium point are negative, the point is stable. The Jacobian of the SIC Model is:
\[
\left( \begin{array}{ccccc}
-(\alpha_1+\alpha_2) & 0 & 0 & 0 & 0 \\
\alpha_1 & -(\delta+\beta) & 0 & 0 & 0 \\
0 & \delta & -(\delta+\beta) & 0 & 0 \\
0 & 0 & \delta & \gamma_1T-\epsilon & \gamma_1A \\
0 & 0 & 0 & \epsilon-\gamma_1T & -\gamma_1A-\gamma_2
\end{array} \right).
\]
\indent Given the block structure of this matrix, the first three eigenvalues are on the diagonal, and we observe that they are all negative. Therefore, isolating the bottom right $2\times2$ matrix, we have:
\begin{center}
$Y=\begin{pmatrix}
\gamma_1T-\epsilon & \gamma_1A \\
\epsilon-\gamma_1T & -\gamma_1A-\gamma_2
\end{pmatrix}$.
\end{center}

In order to have stability, the trace of this matrix must be negative, and the determinant  must be positive, as follows:
\begin{eqnarray}
\mbox{Tr(Y)}&=&\gamma_1T-\epsilon-\gamma_1A-\gamma_2 < 0, \nonumber \\
\mbox{Det(Y)}&=&\epsilon\gamma_2-\gamma_1\gamma_2T > 0. \nonumber
\end{eqnarray}
\indent Substituting the equilibrium points into the trace and determinant yields:
\begin{eqnarray}
\mbox{Tr(Y)}&=&\left(\frac{\gamma_1\delta}{\gamma_2}\right)C_2^*-\epsilon-\frac{\delta\gamma_1C_2^*}{\epsilon-\left(\frac{\delta\gamma_1}{\gamma_2}\right) C_2^*}-\gamma_2 < 0, \nonumber \\
\mbox{Det(Y)}&=&\epsilon\gamma_2-\delta\gamma_1C_2^* > 0.\nonumber
\end{eqnarray}
\indent For Det(Y) $>$ 0, we have,
\begin{eqnarray}
\epsilon - \frac{\delta \gamma_1}{\gamma_2}C_2^* = \gamma_1\left(\frac{\epsilon}{\gamma_1} - \frac{\delta}{\gamma_2}C_2^*\right) = \gamma_1(\frac{\epsilon}{\gamma_1} - T^*) > 0. \nonumber
\end{eqnarray}
\indent Simplifying the condition on the determinant, we obtain:
\begin{eqnarray}
\delta\gamma_1C_2^*&<&\epsilon\gamma_2 \nonumber \\
C_2^*&<&\frac{\epsilon\gamma_2}{\delta\gamma_1} \nonumber \\
T^*&<&\frac{\epsilon}{\gamma_1}. \nonumber
\end{eqnarray}
\indent This is the same condition we had before for the existence of a biologically relevant $A^*$. \\
\begin{figure}[htbp]
\begin{center}
\includegraphics[scale=.5]{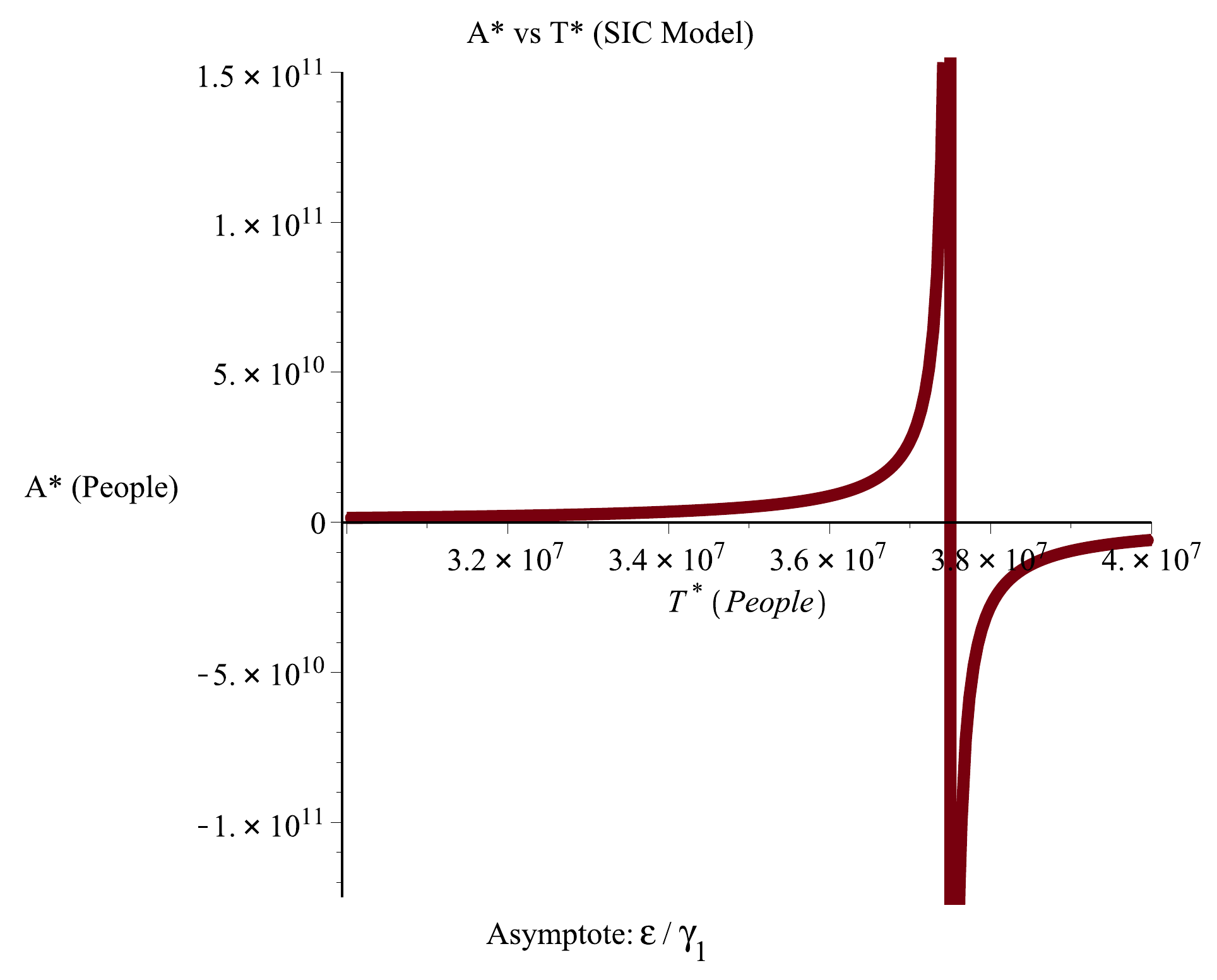} \\
\caption{\small Plot of $T^*$ vs. $A^*$. \smaller{This shows that $A^*$ is positive when $T^*<\frac{\epsilon}{\gamma_1}$. Note that the vertical asymptote occurs at $T^*=\frac{\epsilon}{\gamma_1}$ where $\epsilon=0.0300$ and $\gamma_1=0.0000000008$}}
\label{infinityBifurcation}
\end{center}
\end{figure}
\indent Simplifying the trace: 
\begin{eqnarray}
\mbox{Tr(Y)}&=&\left(\frac{\gamma_1\delta}{\gamma_2}\right)C_2^*\left(\epsilon-\left(\frac{\delta\gamma_1}{\gamma_2}\right) C_2^*\right)-\epsilon\left(\epsilon-\left(\frac{\delta\gamma_1}{\gamma_2}\right) C_2^*\right)-\delta\gamma_1C_2^*-\gamma_2\left(\epsilon-\left(\frac{\delta\gamma_1}{\gamma_2}\right) C_2^*\right) \nonumber \\
&=&-\left(\left(\frac{\gamma_1\delta}{\gamma_2}\right)C_2^*\right)^2+2\epsilon\left(\frac{\gamma_1\delta}{\gamma_2}\right)C_2^*-\epsilon^2-\epsilon\gamma_2 \nonumber \\
&=&-\left(\left(\frac{\gamma_1\delta}{\gamma_2}\right)C_2^*-\epsilon\right)^2-\epsilon\gamma_2 < 0. \nonumber
\end{eqnarray}

Therefore, the trace is always negative, and the equilibrium point needs $T^* <\frac{\epsilon}{\gamma_1}$ in order to be stable. When $T^* > \frac{\epsilon}{\gamma_1}$, there is a saddle and a change in stability. Plotting $A^*$ vs. $T^*$ in Figure \ref{infinityBifurcation}, we see that to the left of $T^*$ = $\frac{\epsilon}{\gamma_1}$, we have $A^* > 0$, and $A^* \rightarrow \infty$ as $T^* \rightarrow \frac{\epsilon}{\gamma_1}$. To the right of $T^* = \frac{\epsilon}{\gamma_1}$, we have $A^* < 0$. The switch in stability appears at $T^* = \frac{\epsilon}{\gamma_1}$, and we thus conclude that we have a bifurcation at infinity. The population, however, is still bounded for $T^*<\frac{\epsilon}{\gamma_1}$ (see Appendix II). \\
\indent Setting parameters to values that fall within the realistic ranges yields a stable equilibrium. When reviewing this model, the issue of a bifurcation at infinity when $T^* = \frac{\epsilon}{\gamma_1}$ and the observation that $A$ appears unbounded for $T^* \geq \frac{\epsilon}{\gamma_1}$ point leads us to modify this non-linear model, as described in the next section. 



\subsection{Social Interaction with Abuse-Dependent Prescription Rate (SIAD) Model}
We consider a model that expresses the entrance rate of new Vicodin patients into the system as an inverse function of the abuser population ($A$). A program in California indicates shows that when prescribers are aware of the risks, the number of Vicodin prescriptions decreases by 95\% \cite{oneill2013}. With this model, we determine how varying the flow of new Vicodin patients into the acute medical user ($M$) compartment affects the total population of abusers ($A$) (see Figure 9).

\begin{figure}[htbp]
\begin{center}
\includegraphics[height=4cm,width=12cm]{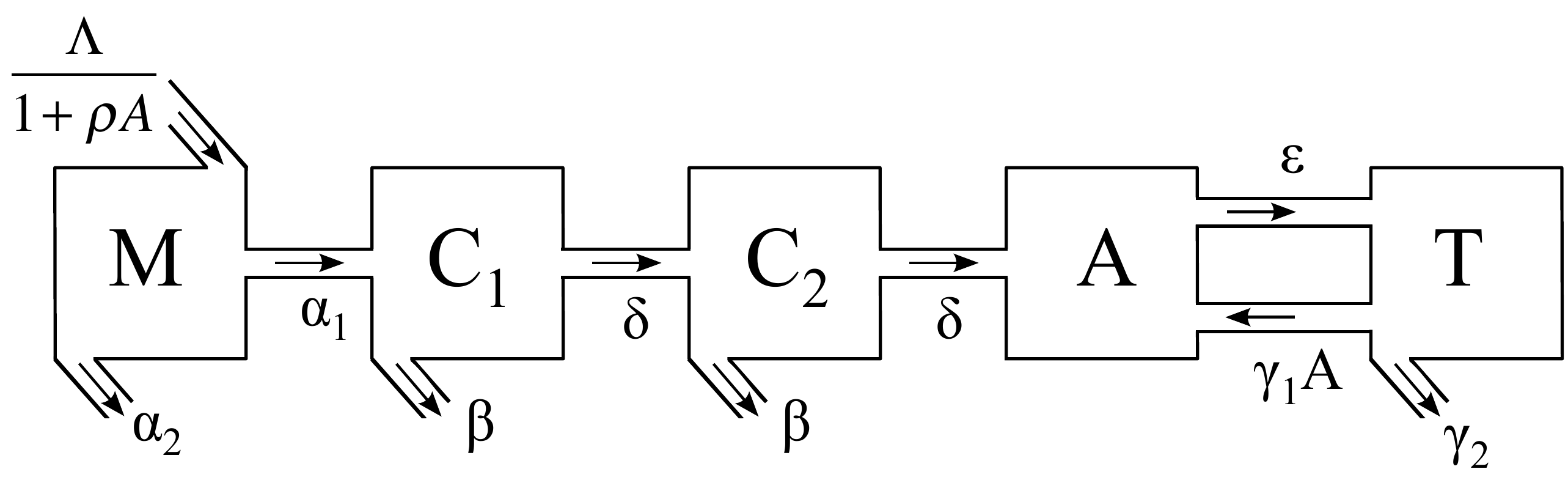} \\
\caption{\small Social Interaction with Abuse-Dependent Prescription Rate (SIAD) Model. \smaller{This model considers $\Lambda$ to be dependent on the population of the ($A$) compartment.}}
\label{rhoModel}
\end{center}
\end{figure}

The governing equations of this non-linear system are given by:
\begin{eqnarray}
\frac{dM}{dt} &=& \frac{\Lambda}{1+ \rho A} - (\alpha_1 + \alpha_2)M \nonumber \\
\frac{dC_1}{dt} &=& \alpha_1 M - (\delta + \beta)C_1 \nonumber \\
\frac{dC_2}{dt} &=& \delta C_1 - (\delta + \beta)C_2 \nonumber \\
\frac{dA}{dt} &=& \gamma_1AT - \epsilon A+ \delta C_2 \nonumber \\
\frac{dT}{dt} &=& -\gamma_1 AT + \epsilon A - \gamma_2T \nonumber
\end{eqnarray}
\indent The parameter $\rho$, which has unit $\frac{1}{\mbox{people}}$, determines the rate at which the entrance rate decreases with respect to the abuser population ($A$). When $\rho$ is small, the system behaves similarly to the SIC Model. When $\rho$ is large, the entrance function decreases quickly.\\
\indent We analyze the stability of the equilibria of this system numerically.There are two equilibrium points, one yields all positive values while the other yields all negative values. We know that there is some type of unstable object between the two equilibrium points. We do not know what this is, but we know it exists in dynamical theory. For our paramater ranges, both equilibria are stable; however, only the positive equilibrium is bioligically relevant. 

\begin{figure}[htbp]
\begin{center}
\includegraphics[scale=0.7]{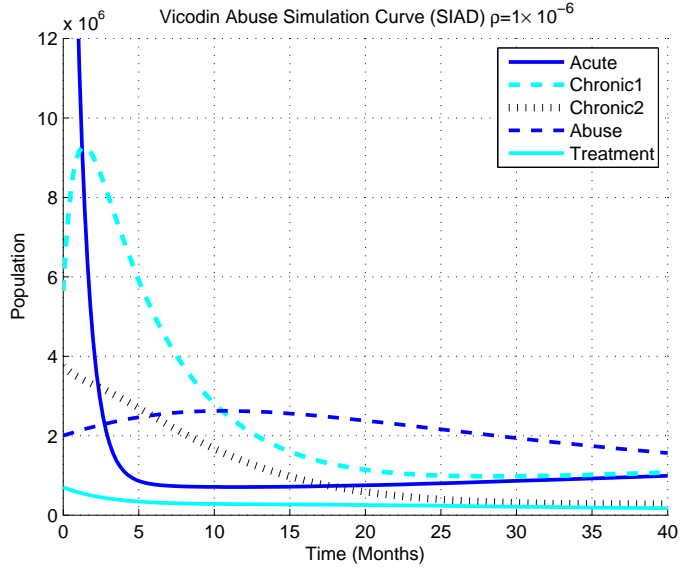} \\
\caption{\small Numerical Solutions of the SIAD Model. \smaller{This plot illustrates the numerical solutions of the SIAD Model. Note that this model has 500,000 fewer abuser after 40 months than in the SIC Model}}
\label{numericalrho}
\end{center}
\end{figure}


\subsection{Adjoint Sensitivity Analysis of the SIAD Model}\label{subsec:adjointrho}
We utilize the adjoint method for sensitivity analysis to examine this model \cite{BradleyPDE}. We focus on small perturbations of each parameter to determine the effect on the population of abusers ($A$). Figure~\ref{nonlinsens} on page \pageref{nonlinsens} demonstrates these findings. Recall the system of equations of the SIAD Model is:
\begin{eqnarray}
\frac{dM}{dt}&=& \frac{\Lambda}{1+\rho A}-(\alpha_1+\alpha_2)M \nonumber \\
\frac{dC_1}{dt}&=&\alpha_1M-(\delta+\beta)C_1 \nonumber \\
\frac{dC_2}{dt}&=&\delta C_1-(\delta+\beta)C_2 \nonumber \\
\frac{dA}{dt}&=&\delta C_2+\gamma_1AT-\epsilon A \nonumber \\
\frac{dT}{dt}&=&\epsilon A-\gamma_1AT-\gamma_2T. \nonumber 
\end{eqnarray}
\indent Let $\dot{M}$ denote $\frac{dM}{dt}$, $\dot{C_1}$ denote $\frac{dC_1}{dt}$, $\dot{C_2}$ denote $\frac{dC_2}{dt}$, $\dot{A}$ denote $\frac{dA}{dt}$, and $\dot{T}$ denote $\frac{dT}{dt}$. Using the adjoint sensitivity method \cite{BradleyPDE}, we rewrite the system of ordinary differential equations: 
\begin{center}
$F(t,$ $\vec{x}$, $\dot\vec{x}$, $\vec{p}) = \begin{bmatrix}
	\dot{M}-\frac{\Lambda}{1+\rho A}+(\alpha_1+\alpha_2)M \\
	\dot{C_1}-\alpha_1M+(\delta+\beta)C_1 \\
	\dot{C_2}-\delta C_1+(\delta+\beta)C_2 \\
	\dot{A}-\delta C_2-\gamma_1AT+\epsilon A \\
	\dot{T}-\epsilon A+\gamma_1AT+\gamma_2T 
\end{bmatrix} = 0$,
\end{center}
\indent where
$\vec{x}^T = \begin{bmatrix}
	M & C_1 & C_2 & A & T
\end{bmatrix}$  is the vector containing the different compartment populations over time, and 
$\vec{x}(0)^T = \begin{bmatrix}
	M(0) & C_1(0) & C_2(0) & A(0) & T(0)
\end{bmatrix}$ is the vector containing the initial population sizes of each compartment. \\
\indent We define our parameter vector $\vec{p}$ such that \setcounter{MaxMatrixCols}{15}
$\vec{p}^T = \begin{bmatrix}
	\Lambda & \rho & \alpha_1 & \alpha_2 & \delta & \beta & \epsilon & \gamma_1 & \gamma_2 & u_1 & u_2 & u_3 & u_4 & u_5
\end{bmatrix}$, where $u_i$ for $i$=1,...,5 represents our initial condition parameters for which the sensitivity index is computed. For the purpose of calculating the sensitivity of our solutions to our initial conditions, we define 
\begin{center}
$\vec{y}(0) = \begin{bmatrix}
	M(0)(1-u_1) \\
	C_1(0)(1-u_2) \\
	C_2(0)(1-u_3) \\
	A(0)(1-u_4) \\
	T(0)(1-u_5)
\end{bmatrix}$.
\end{center}
$\vec{y}$(0) is the vector that contains the $u_i$ (for $i$=1,...,5) percent change of the initial population sizes of the compartments. \\
\indent Because we want to minimize the population of abusers ($A$), we define the objective function $A(\vec{x}, \vec{p}) = \int^T_0 g(\vec{x}, t, \vec{p}) dt = \int^T_0 \dot{A} dt$. 
We analyze this because we are interested in how much the abuser population ($A$) is affected by small changes in the parameters and the initial population of each compartment ($\vec{p}$). \\
\indent Following the second step of the algorithm for computing the sensitivity equations \cite{BradleyPDE}, the adjoint is $g_x + \lambda^T(F_x-\dot{F}_{\dot{x}})-\dot{\lambda}^T F_{\dot{x}} = 0$, where: 
\begin{itemize}
\item $\lambda$ is the Lagrange multiplier, and $\lambda^T = \begin{bmatrix}
		\lambda_1 & \lambda_2 & \lambda_3 & \lambda_4 & \lambda_5
	\end{bmatrix}$.

\item $g_x$ is the partial of $A$ with respect to the population sizes of the compartments ($\vec{x}$):
\begin{center}
$g_x = \frac{\partial\dot{A}}{\partial\vec{x}} = \begin{bmatrix}
	\frac{\partial\dot{A}}{\partial M} & \frac{\partial\dot{A}}{\partial C_1} & \frac{\partial\dot{A}}{\partial C_2} & \frac{\partial\dot{A}}{\partial A} & \frac{\partial\dot{A}}{\partial T}
\end{bmatrix} = \begin{bmatrix}
	0 & 0 & \delta & \gamma_1T-\epsilon & \gamma_1A
\end{bmatrix}$.
\end{center}

\item $F_x$ is the partial of our system of ODEs ($F$) with respect to $\vec{x}$
\begin{center}
$F_x = \begin{bmatrix}
	(\alpha_1+\alpha_2) & 0 & 0 & \frac{\rho\Lambda}{(1+\rho A)^2} & 0 \\
	-\alpha_1 & (\delta+\beta) & 0 & 0 & 0 \\
	0 & -\delta & (\delta+\beta) & 0 & 0 \\
	0 & 0 & -\delta & \epsilon-\gamma_1T & -\gamma_1A \\
	0 & 0 & 0 & \gamma_1T-\epsilon & \gamma_1A+\gamma_2
\end{bmatrix}$.
\end{center}

\item $\dot{\vec{x}}$ is the derivative of the compartment vector ($\vec{x}$) with respect to time $\left(\dot{\vec{x}}^T = \begin{bmatrix}
	\dot{M} & \dot{C_1} & \dot{C_2} & \dot{A} & \dot{T}
\end{bmatrix}\right)$.

\item $F_{\dot{x}}$ is the partial of our system of ODEs ($F$) with respect to the $\dot{\vec{x}}$:
\begin{center}
$F_{\dot{x}} = \begin{bmatrix}
	1 & 0 & 0 & 0 & 0 \\
	0 & 1 & 0 & 0 & 0 \\
	0 & 0 & 1 & 0 & 0 \\
	0 & 0 & 0 & 1 & 0 \\
	0 & 0 & 0 & 0 & 1
\end{bmatrix}$.
\end{center}

\item $\dot{F}_{\dot{x}}$ is the partial of the derivative of our ODE system ($\dot{F}$) with respect to $\dot{\vec{x}}$ and $\dot{F}_{\dot{x}} = 0_{5\times5}$.
\end{itemize}

\indent Now, solving for the $\lambda^T(F_x-\dot{F}_{\dot{x}})$ part of the adjoint, we note that $F_x-\dot{F}_{\dot{x}} = F_x$. Therefore, multliplying the transpose of the Lagrange multiplier vector by the partial of our system of ODEs with respect to the compartmental populations yields 
$\lambda^T F_x = \begin{bmatrix} v & w \end{bmatrix}$, where
\begin{eqnarray}
v &=& \begin{bmatrix}(\alpha_1+\alpha_2)\lambda_1-\alpha_1\lambda_2 & (\delta+\beta)\lambda_2-\delta\lambda_3\ & (\delta+\beta)\lambda_3-\delta\lambda_4 \end{bmatrix}\nonumber \\
w &=& \begin{bmatrix} \frac{\rho\Lambda}{(1+\rho A)^2}\lambda_1+\epsilon(\lambda_4-\lambda_5)+T\gamma_1(\lambda_5-\lambda_4) & -\gamma_1A\lambda_4+(\gamma_1A+\gamma_2)\lambda_5 \end{bmatrix}.\nonumber
\end{eqnarray}
\indent For the final section of the adjoint ($\dot{\lambda}^T F_{\dot{x}}$), the product of the transpose of the Lagrange multiplier vector and the partial of our system of ODEs with respect to the derivative of the compartmental populations is
$\dot{\lambda}^T F_{\dot{x}} = \begin{bmatrix}
	\dot{\lambda_1} & \dot{\lambda_2} & \dot{\lambda_3} & \dot{\lambda_4} & \dot{\lambda_5}
\end{bmatrix}$. \\
\indent Combining the results from above, we obtain the adjoint equation:
\begin{eqnarray}
(\alpha_1+\alpha_2)\lambda_1-\alpha_1\lambda_2-\dot{\lambda}_1&=&0 \nonumber \\
(\delta+\beta)\lambda_2-\delta\lambda_3-\dot{\lambda}_2&=&0 \nonumber \\
\delta+(\delta+\beta)\lambda_3-\delta\lambda_4-\dot{\lambda}_3&=&0 \nonumber \\
\left(\frac{\rho\Lambda}{(1+\rho A)^2}\right)\lambda_1+(\epsilon-\gamma_1T)(\lambda_4-\lambda_5 -1)-\dot{\lambda}_4&=&0 \nonumber\\
\gamma_1A(1-\lambda_4) +(\gamma_1A+\gamma_2)\lambda_5-\dot{\lambda}_5&=&0 \nonumber
\end{eqnarray}
with initial conditions $\lambda_i$(T)=0, for $i$ = 1, ..., 5. \\
\indent Now, we simultaneously solve the initial value problem: $F=0$, \\$\vec{x}(0)^T = \begin{bmatrix} M(0) & C_1(0) & C_2(0) & A(0) & T(0) \end{bmatrix}$. Also we solve the general sensitivity equation, 
$\frac{dA}{dp}=\int_0^T (g_p+\lambda^TF_p) dt+\lambda^T F_{\dot{x}}|_{t=0} \vec{y}_{x(0)}^{~-1}\vec{y_p}$, where:
\begin{itemize}
\item $g_p$ is the partial of $\dot{A}$ with respect the the vector of parameters ($\vec{p}$)
\begin{center}
$g_p = \begin{bmatrix}
	0 & 0 & 0 & 0 & C_2 & 0 & -A & AT & 0 & 0 & 0 & 0 & 0 & 0
\end{bmatrix}$.
\end{center}

\item $F_p$ is the partial of our system of ODEs ($F$) with respect to the the parameters ($\vec{p}$):
\begin{center}
$F_p=\begin{bmatrix}
-\frac{1}{1+\rho A}&\frac{\Lambda A}{(1+\rho A)^2}&M&M&0&0&0&0&0& & & & &\\
0&0&-M&0&C_1&C_1&0&0&0& & & & & \\ 
0&0&0&0&C_2-C_1&C_2&0&0&0 & &\bigzero& & &\\ 
0&0&0&0&-C_2&0&A&-AT & 0 & & & 5\times5 & &\\ 
0&0&0&0&0&0&-A&AT&T & & & & &
\end{bmatrix}$.
\end{center}

\item $\vec{y}_{p}$ is the partial of the percent change in initial compartment populations ($\vec{y}$) with respect to the parameters ($\vec{p}$):
\begin{center}
$\vec{y}_{p}=\begin{bmatrix} &&&&&&&-M(0)&0&0&0&0\\&&&&&&&0&-C_1(0)&0&0&0\\&&&\bigzero&&&&0&0&-C_2&0&0\\&&&&5\times8&&&0&0&0&-A(0)&0\\&&&&&&&0&0&0&0&-T(0) \end{bmatrix}$.
\end{center}

\item $\vec{y}_{x(0)}^{~-1}$ is the inverse of the partial of the percent change in initial compartment populations ($\vec{y}$) with respect to the initial population vector ($\vec{x}$(0)): \\
$\vec{y}_{x(0)}=\begin{bmatrix} 1-u_1&0&0&0&0 \\ 0&1-u_2&0&0&0 \\ 0&0&1-u_3&0&0 \\ 0&0&0&1-u_4&0 \\ 0&0&0&0&1-u_5 \end{bmatrix}$,~
$\vec{y}_{x(0)}^{~-1}=\begin{bmatrix} \frac{1}{1-u_1}&0&0&0&0 \\ 0&\frac{1}{1-u_2}&0&0&0 \\ 0&0&\frac{1}{1-u_3}&0&0 \\ 0&0&0&\frac{1}{1-u_4}&0 \\ 0&0&0&0&\frac{1}{1-u_5} \end{bmatrix}$.

\item All previously seen expressions ($\lambda^T$, $F_{\dot{x}}$) have the same definitions from above.
\end{itemize}

\indent Within the integrand of the general sensitivity equation, we have the product of the Lagrange multiplier vector ($\lambda^T$) and the partial derivative of our system of ODEs with respect to the the parameters ($F_p$), which is:
\begin{center}
$\lambda^T F_p = \begin{bmatrix}v & w\end{bmatrix}$, where
\begin{eqnarray}
v &=&\begin{bmatrix}-\frac{\lambda_1}{1+\rho A} & \frac{\lambda_1\Lambda A}{(1+\rho A)^2} &(\lambda_1-\lambda_2) M & \lambda_1M & (\lambda_2-\lambda_3)C_1+(\lambda_3-\lambda_4)C_2\end{bmatrix}\nonumber \\
w &=&\begin{bmatrix} \lambda_2C_1+\lambda_3C_2 & (\lambda_4-\lambda_5)A & (-\lambda_4+\lambda_5)AT & \lambda_5 T & 0 & 0 & 0 & 0 & 0\end{bmatrix}.\nonumber
\end{eqnarray}
\end{center}

\indent Multiplying the the Lagrange multiplier vector ($\lambda^T$) by the partial of our system of ODEs with respect to the the derivative of the compartment vector with respect to time ($F_{\dot{x}}$) with the inverse of the partial of the percent change in initial compartment populations with respect to the initial population vector ($\vec{y}_{x(0)}^{~-1}$), and with the partial of the percent change in initial compartment populations with respect to the parameters ($\vec{y}_{p}$), we have: 
\begin{center}
$F_{\dot{x}}|_{t=0} \vec{y}_{x(0)}^{~-1}\vec{y_p}=
\begin{bmatrix} &&&&&&&&&\frac{-M(0)}{1-u_1}&0&0&0&0\\&&&&&&&&&0&\frac{-C_1(0)}{1-u_2}&0&0&0 \\ &&&\bigzero&&&&&&0&0&\frac{-C_2(0)}{1-u_3}&0&0 \\ &&&&5\times9&&&&&0&0&0&\frac{-A(0)}{1-u_4}&0 \\ &&&&&&&&&0&0&0&0&\frac{-T(0)}{1-u_5} \end{bmatrix}$,
\end{center}
$\lambda^TF_{\dot{x}}|_{t=0} \vec{y}_{x(0)}^{~-1}\vec{y_p} = \begin{bmatrix}
	0 & 0 & 0 & 0 & 0 & 0 & 0 & 0 & 0 & -\lambda_1\frac{M(0)}{1-u_1} & -\lambda_2\frac{C_1(0)}{1-u_2} & -\lambda_3\frac{C_2(0)}{1-u_3} & -\lambda_4\frac{A(0)}{1-u_4} & -\lambda_5\frac{T(0)}{1-u_5}
\end{bmatrix}$.\\
\indent Thus, from $\frac{dA}{dp}$, we add $g_p$ and $\lambda^T F_p$ together within the integrand and then add $\lambda^TF_{\dot{x}}|_{t=0} \vec{y}_{x(0)}^{~-1}\vec{y_p}$ outside of it. Now we have the sensitivity equations:
\begin{eqnarray}
\frac{\partial A}{\partial \Lambda}&=&\int_0^T -\frac{\lambda_1}{1+\rho A}dt \nonumber\\
\frac{\partial A}{\partial \rho}&=&\int_0^T \frac{\lambda \Lambda A}{(1+\rho A)^2}dt \nonumber\\
\frac{\partial A}{\partial \alpha_1} &=& \int_0^T (\lambda_1-\lambda_2) M dt \nonumber\\
\frac{\partial A}{\partial \alpha_2} &=& \int_0^T \lambda_1Mdt \nonumber\\
\frac{\partial A}{\partial \delta} &=& \int_0^T (\lambda_2-\lambda_3)C_1+(\lambda_3-\lambda_4+1)C_2dt \nonumber\\
\frac{\partial A}{\partial \beta} &=& \int_0^T \lambda_2C_1+\lambda_3C_2 dt\nonumber\\
\frac{\partial A}{\partial \epsilon} &=& \int_0^T (\lambda_4-\lambda_5-1)A dt \nonumber\\
\frac{\partial A}{\partial \gamma_1} &=& \int_0^T (-\lambda_4+\lambda_5+1)AT dt \nonumber \\
\frac{\partial A}{\partial \gamma_2} &=& \int_0^T \lambda_5Tdt \nonumber\\
\frac{\partial A}{\partial u_1} &=& -\lambda_1\frac{M(0)}{1-u_1} \nonumber\\
\frac{\partial A}{\partial u_2} &=& -\lambda_2\frac{C_1(0)}{1-u_2} \nonumber\\
\frac{\partial A}{\partial u_3} &=& -\lambda_3\frac{C_2(0)}{1-u_3} \nonumber\\
\frac{\partial A}{\partial u_4} &=& -\lambda_4\frac{A(0)}{1-u_4} \nonumber\\
\frac{\partial A}{\partial u_5} &=& -\lambda_5\frac{T(0)}{1-u_5}. \nonumber
\end{eqnarray}
\indent By varying the upper limit of integration, T, from [0, 60], we are able to get the sensitivity of the population of the abusers ($A$) with respect to each parameter over the first 60 months using MATLAB. See Figure~\ref{nonlinsens}.


\subsection{Results and Conclusions of SIAD Model}\label{subsec:resultsrho}
\begin{figure}[http]
\begin{centering}
\includegraphics[scale=0.81]{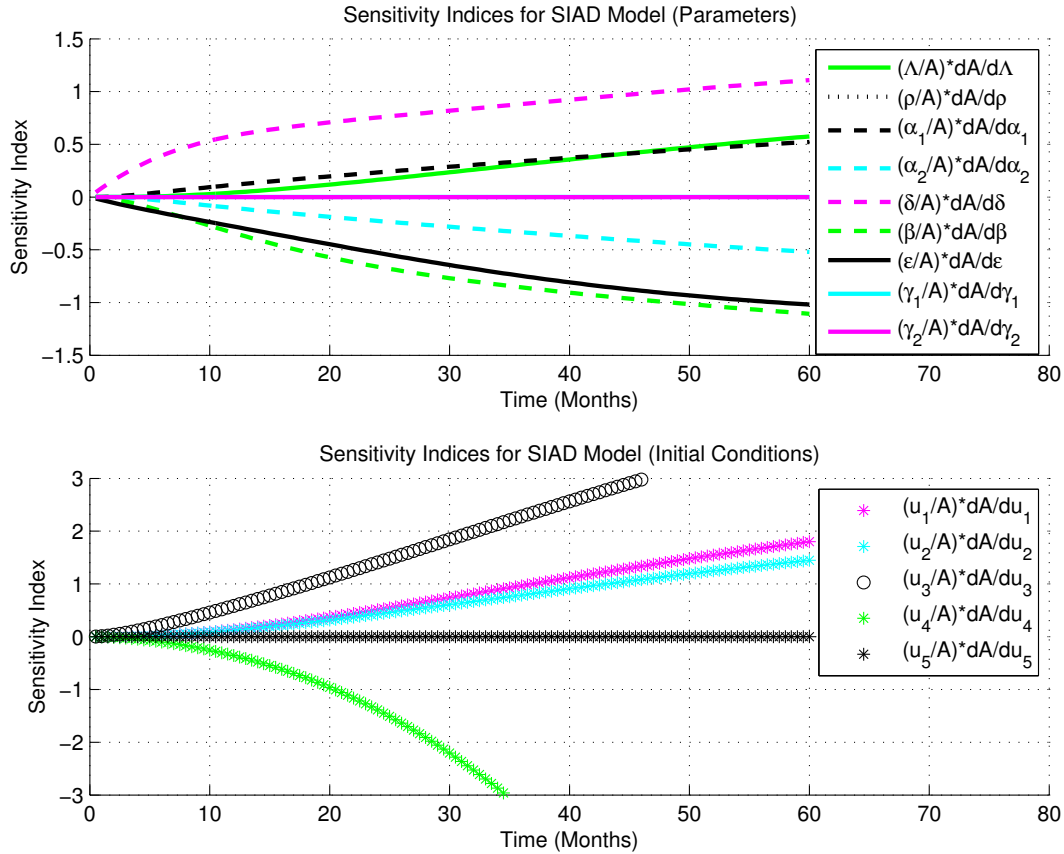}
\caption{\small Sensitivity of the SIAD Model. \smaller{These figures illustrate the sensitivity indices ($\frac{\Lambda}{A} \frac{\partial A}{\partial \Lambda}$, etc.) and indicate the degree to which the parameters affect the number of people in $A$. The top graph indicates that prevention rates ($\delta$ and $\beta$) and the rate at which abusers seek treatment ($\epsilon$) have the greatest influence on the size of the $A$ compartment. The bottom graph indicates that the initial size of the abuser compartment affects the number of abusers the most. (Note: in the top graph $\frac{\alpha_1}{A} \frac{\partial A}{\partial \alpha_1}$ and $\frac{\Lambda}{A} \frac{\partial A}{\partial \Lambda}$ overlap and $\frac{\rho}{A} \frac{\partial A}{\partial \rho}$, $\frac{\gamma_1}{A} \frac{\partial A}{\partial \gamma_1}$, and $\frac{\gamma_2}{A} \frac{\partial A}{\partial \gamma_2}$ are 0)}}
\label{nonlinsens}
\end{centering}
\end{figure}

\begin{figure} [htbp]
\begin{center}
\includegraphics[scale=.8]{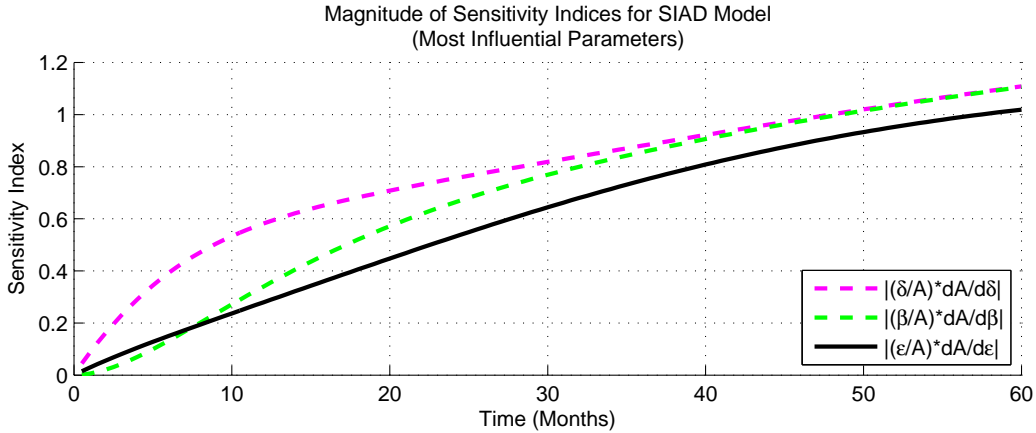}
\end{center}
\caption{\small Sensitivity Index Magnitude Comparison for the SIAD Model. \smaller{This plot compares the magnitudes of the sensitivity indices of the most influential parameters for the SIAD Model from Figure \ref{nonlinsens}. Starting with the strongest influence on the size of the abuser compartment, we have the rate at which chronic users become abusers ($\delta$), the rate at which chronic users stop taking Vicodin ($\beta$), and the rate at which abusers enter treatment ($\epsilon$). However, after 50 months have passed, $\delta$ and $\beta$ have the same influence on $A$. The sensitivity indices of the initial conditions were not considered, because the current population values cannot be changed.}}
\label{nonlinsensmag}
\end{figure}

Figure~\ref{nonlinsens} and \ref{nonlinsensmag} imply that the rate of transition to abuse from chronic medical use of Vicodin ($\delta$) has a strong positive correlation with the abuser population ($A$). Additionally, the rates of movement from abuse to treatment ($\epsilon$) and movement out of the population from chronic medical use ($\beta$) have strong negative correlations. We conclude that the rate at which chronic medical users of Vicodin become abusers has the greatest influence on the total number of abusers both initially and as time progresses. The rate at which Vicodin abusers enter treatment has a strong inverse relation to the size of the abuser compartment. That is, increases in the rate of abusers seeking treatment has a large impact on reducing the number of abusers. The rate at which chronic medical users stop taking Vicodin has a stronger negative correlation than the rate at which abusers enter treatment. Note that intially $\delta$ has the greatest influence on $A$, but after approximately 60 months, the magnitudes of influence of $\epsilon$ and $\beta$ near that of $\delta$.\\ 
\indent The initial sizes of the chronic medical user compartments ($C_2$) have positive correlations with the size of the abuser compartment ($A$). This makes sense in the context of our model, because the only pathway to abuse is through chronic medical use of Vicodin. Additionally, all other compartments, with the exception of the initial population of the abuser compartment, have positive correlations with the size of the abuser population. This enables more people to have a potential for drug abuse of Vicodin. The initial size of the abuser compartment has a negative correlation because there is an inverse relationship between the number of abusers and the rate of new Vicodin-prescribed patients. Thus, the larger the initial abuser compartment is, the smaller the rate at which newly prescribed medical users enter the population. A large initial number of Vicodin abusers causes a decrease in the abuser compartment. \\

\indent Analyzing the sensitivity of the SIAD Model, we determine that $\delta$ and $\beta$ have the greatest influence on the abuser population ($A)$, while $\gamma_1$ and $\gamma_2$ have no influence. Because $\delta$ and $\beta$ are associated with prevention and $\gamma_1$ and $\gamma_2$ are associated with treatment, we conclude that improving prevention is the most effective strategy for addressing the Vicodin abuse problem when considering this model. 


\section{Results and Conclusions}

The CVT model assumes no interaction between any compartmental populations. After conducting the linear analysis, two additional models, one with a single non-linear term (SIC) and the other with two (SIAD), were analyzed. Placing a dependence on the abusers ($A$) for the relapse rate introduces a social aspect to the model.  Thus, the number of people in the $A$ compartment increases, individuals in treatment have more contact with abusers and are more prone to relapse. In the SIAD model, we introduce a second non-linear term, an inverse relationshipe between the number of abusers and the number of new Vicodin patients. This suggests that as the abusive compartment grows, doctors and/or patients become aware of this, and less Vicodin is prescribed. These changes add more realistic dimensions to the model, as these social interactions do take place and have an influence on the flow within the model. Conducting sensitivity analyses on these models, we consider prevention to be the most effective method of controlling Vicodin abuse in a population that considers only those who are initially given a prescription for the drug.\\
\indent All three models all indicate that the parameters representing exits from population via the chronic compartments ($\beta$) and entrance into the abuser compartment from $C_2$ ($\delta$) have the greatest impact on the population of abusers ($A$). Relative to other factors, changes in treatment success ($\gamma_2$) and failure ($\gamma_1$) have little effect on the number of Vicodin abusers. For example, the magnitude of the sensitivity index for $\delta$ is nearly three times larger than that of $\gamma_1$ in the CVT Model. In the SIAD Model, the magnitudes of the sensitivity indices for $\delta$ and $\beta$ are greater than in the CVT model, while the indices for $\gamma_1$ and $\gamma_2$ are $0$. This means that in the SIAD model, changes in rates of relapse and successful treatment have no effect on the number of abusers. Additionally, in the SIAD Model, the importance of intervention (displayed by our parameter connecting the abuser compartment to the treatment compartment, $\epsilon$) influences the number of abusers. In the short term, $\delta$ has a greater impact on $A$, while in the long term, $\delta$ and $\beta$ have similar influences. Because both of $\delta$ and $\beta$ are associated with prevention rather than treatment, we determine that whether focusing on the short term or the long term, prevention measures are more effective and should be the focus of controlling the Vicodin abuse problem.\\
\indent Fluctuations in these prevention parameters have a more significant effect on the number of abusers over time. More specifically, implementing prevention measures that lower the rate of chronic medical users becoming abusers ($\delta$) or that raise the rate at which chronic patients stop taking Vicodin ($\beta$) lowers the abuser population ($A$) in a more significant manner than implementing treatment programs to lower the relapse rate ($\gamma_1$) and raise the successful treatment rate ($\gamma_2$).\\
\indent While these models give insight into the Vicodin abuse problem, limitations exist. Many of our parameters are not independent. For example, the value of the relapse rate ($\gamma_1$) depends upon the successful treatment rate ($\gamma_2$). For additional parameter value calculations, see Appendix IV. In these analyses, we assume that only one parameter value varies at a time, which may not be feasible. Additionally, the bifurcation at infinity in the SIC Model creates a situation in which the model is difficult to analyze in a biologically relevant manner. Furthermore, the acute medical user population ($M$) decreases dramatically in the first few months of each model, and that behavior does not seem consistent with the data \cite{mccabe2005illicit, CDCMorbidityOnPill, edlund2010risks, CDCPolicyImpact, sung2005nonmedical}. For the purposes of this model, we assume parameters to be constant over time. It is likely, however, that these parameter values change over time.

\subsection{Future Work}

Our models could be further adapted to include those Vicodin abusers who were not introduced to the drug via prescription. Also, more non-linear terms depicting realistic interactions could give better results, or parameter change over time could be considered instead of assuming constant parameter values over time. Also, different restrictions could be placed on the models so that they depict only certain demographics. \\
\indent Another possible future research pathway is to consider an economic application. Cost analysis could be incorporated in order to determine the most cost-effective method to reduce the population of abusers. We could also model the flow of Vicodin pills from the manufacturer to abusers.


\section{Acknowledgments}
We would like to acknowledge Dr. Carlos Castillo-Chavez, Executive Director for the Mathematical and Theoretical Biology Institute (MTBI), for his continued support and commitment to fostering academic development and providing opportunities for intellectual growth. We also wish to extend gratitude to Dr. Erika T. Camacho and Dr. Stephen A. Wirkus, Co-Executive Summer Directors of MTBI, for their ongoing guidance throughout this program. We extend thanks to Yiqiang Zheng for his advice and help as well. Additionally, we would like to thank all the faculty, students, and staff of MTBI for their willingness to share their knowledge and expertise.\\
\indent This research was conducted at MTBI at Arizona State University. This program is partially supported by grants from the National Science Foundation (NSF - Grant DMPS-1263374), the National Security Agency (NSA - Grant H98230-13-1-0261), the Office of the President of ASU, and the Office of the Provost of ASU. 
\newpage


\section*{Appendix I}
To determine stability, we examine the eigenvalues of our system:
\vspace{-.2in}
\begin{eqnarray}
\lambda_1 &=& - (\alpha_1 + \alpha_2) \nonumber \\
\lambda_3  ~=~ \lambda_2 &=& - (\delta + \beta) \nonumber \\
\lambda_4 &=& \frac{- (\epsilon + \gamma_1 + \gamma_2) + \sqrt{\epsilon^2 + 2(\gamma_1 - \gamma_2)\epsilon + (\gamma_1 + \gamma_2)^2}}{2} \nonumber \\
\lambda_5 &=& \frac{- (\epsilon + \gamma_1 + \gamma_2) - \sqrt{\epsilon^2 + 2(\gamma_1 - \gamma_2)\epsilon + (\gamma_1 + \gamma_2)^2}}{2} \nonumber
\end{eqnarray}
Because $\alpha_1$, $\alpha_2$, $\beta$, $\delta$ are positive, $\lambda_1$, $\lambda_2$, $\lambda_3$, and $\lambda_5$ are negative. To verify that $\lambda_4$ is negative, we need:
\vspace{-.2in}
\begin{eqnarray}
(\epsilon + \gamma_1 + \gamma_2) &>& \sqrt{\epsilon^2 + 2(\gamma_1 - \gamma_2)\epsilon + (\gamma_1 + \gamma_2)^2} \nonumber \\
\vspace{0.1in}
\Rightarrow [(\epsilon + \gamma_1 + \gamma_2)]^2 &>& \epsilon^2 + 2(\gamma_1 - \gamma_2)\epsilon + (\gamma_1 + \gamma_2)^2 \nonumber \\
\vspace{0.1in}
\Rightarrow \epsilon^2 + 2(\gamma_1 + \gamma_2)\epsilon + (\gamma_1 + \gamma_2)^2 &>& \epsilon^2 + 2(\gamma_1 - \gamma_2)\epsilon + (\gamma_1 + \gamma_2)^2 \nonumber \\
\vspace{0.1in}
\Rightarrow 2(\gamma_1 + \gamma_2)\epsilon &>& 2(\gamma_1 - \gamma_2)\epsilon \nonumber \\
\vspace{0.1in}
\Rightarrow \gamma_1 + \gamma_2 &>& \gamma_1 - \gamma_2 \nonumber \\
\vspace{0.1in}
\Rightarrow \gamma_2 &>& -\gamma_2 \nonumber 
\end{eqnarray}
Because $\gamma_2 > 0$, this statement is always true. Therefore, all eigenvalues are negative, indicating global stability since the system is linear.



\section*{Appendix II}
Claim: the total population $N$ is bounded in the SIC Model for $T^{*}<\frac{\epsilon}{\gamma_{1}}$.
\begin{proof}
If $T^{*}<\frac{\epsilon}{\gamma_{1}}$, then the explicit solution to $\frac{dA}{dt}$, $A(t)=s_{2}-s_{3}e^{-\rho_{3} t}-s_{4}e^{-\rho_{4}t}+ce^{(\gamma_{1}T^{*}-\epsilon)t}$ , is of exponential order, which means that $\exists$ a constant $\rho$ and positive constants $t_{0}$ and $W$ such that
\begin{equation}
\label{eq:bindingEq}
e^{-\rho t}|A(t)|<W \nonumber
\end{equation}
for all $t>t_0$ at which $A(t)$ is defined.\\
For any $\rho>0$ and $\rho_{3}=\delta+\gamma>0,\rho_{4}=\alpha_{1}+\alpha_{2}>0$
\begin{eqnarray}
\space \lim_{t \to \infty}e^{-\rho t}A(t)&=&\lim_{t \to \infty}e^{-\rho t}(s_{2}-s_{3}e^{-\rho_{3} t}-s_{4}e^{-\rho_{4}t}+ce^{(\gamma_{1}T^{*}-\epsilon)t})\nonumber\\
&=&\lim_{t \to \infty}s_{2}e^{-\rho t}-s_{3}e^{-t(\rho +\rho_{3} )}-s_{4}e^{-t(\rho +\rho_{4})}+ce^{-t(\rho -(\gamma_{1}T^{*}-\epsilon))})\nonumber\\
&=& 0\nonumber
\end{eqnarray}
where $T^{*}<\frac{\epsilon}{\gamma_{1}}$. This means that there exists a $W>0$ and $t_{0}>0$ so that $e^{-\rho t}|A(t)|<W$ for $t>t_{0}$. So $A(t)$ is bounded.
\\
\\
Let $N(t)=M(t)+C_{1}(t)+C_{2}(t)+A(t)+T(t)$,
\begin{eqnarray}
\frac{dN}{dt}&=&\Lambda-(\alpha_{2}M+\beta C_{1}+\beta C_{2}+\gamma_{2}T)\nonumber\\\nonumber
&\leq & \Lambda-(\bar{\alpha}M+\bar{\alpha} C_{1}+\bar{\alpha} C_{2}+\bar{\alpha}T)\\
&&\mbox{where } \bar{\alpha}=\mbox{min}( \alpha_{2},\beta,\gamma_{2})\nonumber\\
&=&\Lambda-\bar{\alpha}(M+ C_{1}+ C_{2}+T)\nonumber\\
&=&\Lambda-\bar{\alpha}(N-A)\nonumber\\
&\Rightarrow &\frac{dN}{dt}\leq\Lambda-\bar{\alpha}(N-A)\nonumber
\end{eqnarray}
Using the integrating factor technique with factor $e^{\bar{\alpha}t}$ and exponential order property (\ref{eq:bindingEq}) we can get a bound on $N(t)$.
Because we know  $\frac{dN}{dt}+\bar{\alpha}N\leq \Lambda-\bar{\alpha}A$, we have
\begin{eqnarray}
e^{\bar{\alpha}t}\frac{dN}{dt}+\bar{\alpha}e^{\bar{\alpha}t}N &\leq& (\Lambda-\bar{\alpha}A)e^{\bar{\alpha}t} \nonumber \\
\frac{d}{dt}[Ne^{\bar{\alpha}t}] &\leq& (\Lambda-\bar{\alpha}A)e^{\bar{\alpha}t}\nonumber\\
\int\frac{d}{dt}[Ne^{\bar{\alpha}t}]dt &\leq& \int(\Lambda-\bar{\alpha}A)e^{\bar{\alpha}t}dt\nonumber\\
Ne^{\bar{\alpha}t} &\leq& \frac{\Lambda}{\bar{\alpha}}e^{\bar{\alpha}t}  -\bar{\alpha}\int Ae^{\bar{\alpha}t}dt \nonumber\\
N &\leq& \frac{\Lambda}{\bar{\alpha}}  -\bar{\alpha}e^{-\bar{\alpha}t}\int Ae^{\bar{\alpha}t}dt.\nonumber
\end{eqnarray}
Because $A(t)$ is of exponential order\cite{NotEvilSteve}, as is any exponential function ( $e^{-\bar{\alpha}t}$ is also of exponential order), there exist $W>0$ and $t>t_0$ such that $e^{-\rho t}|A(t)e^{-\bar{\alpha}t}|<W$ for all $t>t_0$, where A(t) is defined.
\begin{eqnarray}
\Rightarrow |A(t)e^{-\bar{\alpha}t}|<We^{-\rho t}\ \mbox{~for any~} \rho>0 \nonumber
\end{eqnarray}
Let $0<\rho<\bar{\alpha}$. Thus, 
\begin{eqnarray}
N(t)&\leq& \frac{\Lambda}{\bar{\alpha}}+\bar{\alpha}e^{-\bar{\alpha t}}\int |A(t)e^{\bar{\alpha} t}|dt\nonumber\\\nonumber
&\leq&\frac{\Lambda}{\bar{\alpha}}+\bar{\alpha}e^{-\bar{\alpha t}}\int |We^{\rho t}|dt\\\nonumber
&=& \frac{\Lambda}{\bar{\alpha}}+\frac{\bar{\alpha}}{\rho}e^{-\bar{\alpha t}}e^{\rho t}\nonumber
\end{eqnarray}
In the limit as $t\rightarrow \infty$, $N*$ is bounded by $\frac{\Lambda}{\bar{\alpha}}$ because $e^{(-\bar{\alpha}+\rho)t}\rightarrow 0$. Therefore, $N(t)$ is constant if $T^{*}<\frac{\epsilon}{\gamma_{1}}$ (and, in particular, $\lim _{t\rightarrow \infty}N(t)=\frac{\Lambda}{\bar{\alpha}}$).
\end{proof}


\section*{Appendix III}
Forward sensitivity analysis of this non-linear model would require simultaneous integration of numerous equations. Adjoint sensitivity allows us to single out specific variables to analyze their sensitivities. Here we consider a simple $SI$ model and show that forward sensitivity analysis and adjoint sensitivity analysis yield the same result. The system of equations of the linear model is:
\begin{eqnarray}
\frac{dS}{dt}&=&\Lambda-\alpha IS - \mu S \nonumber \\
\frac{dI}{dt}&=& \alpha IS - \mu I \nonumber
\end{eqnarray}

\subsection*{Forward Sensitivity Analysis for the SI Model}
Utilizing the simple $SI$ model, we calculate the forward sensitivity equations. They are:
\begin{eqnarray}
\frac{\partial S}{\partial \Lambda} &=& 1 -  \alpha \frac{\partial I}{\partial \Lambda}S - \alpha I \frac{\partial S}{\partial \Lambda}  - \mu \frac{\partial S}{\partial \Lambda} \nonumber \\
\frac{\partial S}{\partial \alpha} &=& -IS - \alpha \frac{\partial I}{\partial \alpha}S - \alpha I \frac{\partial S}{\partial \alpha} - \mu \frac{\partial S}{\partial \alpha} \nonumber \\
\frac{\partial S}{\partial \mu} &=& -\alpha \frac{\partial I}{\partial \mu} - \alpha I \frac{\partial S}{\partial \mu} - S - \mu \frac{\partial S}{\partial \mu} \nonumber \\
\frac{\partial I}{\partial \Lambda} &=& \alpha \frac{\partial I}{\partial \Lambda}S + \alpha I \frac{\partial S}{\partial \Lambda} - \mu \frac{\partial I}{\partial \Lambda} \nonumber \\
\frac{\partial I}{\partial \alpha} &=& IS + \alpha \frac{\partial I}{\partial \alpha}S + \alpha I \frac{\partial S}{\partial \alpha} - \mu \frac{\partial I}{\partial \alpha} \nonumber \\
\frac{\partial I}{\partial \mu} &=&  \alpha \frac{\partial I}{\partial \mu}S + \alpha I \frac{\partial S}{\partial \mu} - I - \mu \frac{\partial I}{\partial \mu} \nonumber \\
\frac{\partial S}{\partial u_1} &=& -\alpha \frac{\partial I}{\partial u_1}S - \alpha I \frac{\partial S}{\partial u_1} - \mu \frac{\partial S}{\partial u_1} \nonumber \\
\frac{\partial S}{\partial u_2} &=& -\alpha \frac{\partial I}{\partial u_2}S - \alpha I \frac{\partial S}{\partial u_2} - \mu \frac{\partial S}{\partial u_2} \nonumber \\
\frac{\partial I}{\partial u_1} &=& \alpha \frac{\partial I}{\partial u_1}S + \alpha I \frac{\partial S}{\partial u_1} - \mu \frac{\partial I}{\partial u_1} \nonumber \\
\frac{\partial I}{\partial u_2} &=& \alpha \frac{\partial I}{\partial u_2}S + \alpha I \frac{\partial S}{\partial u_2} - \mu \frac{\partial I}{\partial u_2} \nonumber
\end{eqnarray}

\subsection*{Adjoint Sensitivity Analysis of the SI Model}
Let $\dot{S}$ denote $\frac{dS}{dt}$ and $\dot{I}$ denote $\frac{dI}{dt}$. Using the adjoint sensitivity method \cite{BradleyPDE}, we rewrite the system of ordinary differential equations: 
\begin{center}
$F(t,$ $\vec{x}$, $\dot\vec{x}$, $\vec{p}) = \begin{bmatrix}
	\dot{S} - \Lambda+\alpha IS + \mu S \\
	\dot{I} -  \alpha IS + \mu I 
\end{bmatrix} = 0$
\end{center}
 where
\vspace{0.1in}
$\vec{x}^T = \begin{bmatrix}
	S & I
\end{bmatrix}$ is the vector containing the different compartment populations over time, and 
$\vec{x}(0)^T = \begin{bmatrix}
	S(0) & I(0) 
\end{bmatrix}$ is the vector containing the initial population sizes of each compartment. \\
We define our parameter vector $\vec{p}$ such that \setcounter{MaxMatrixCols}{14}
$\vec{p}^T = \begin{bmatrix}
	\Lambda & \alpha & \mu & u_1 &  u_2
\end{bmatrix}$, where $u_i$ for $i$=1,2 represents our initial condition parameters for which the sensitivity index will be computed. For the purpose of calculating the sensitivity of our solutions to our initial conditions, we define
\begin{center}
$\vec{y}(0) = \begin{bmatrix}
	S(0)(1-u_1) \\
	I(0)(1-u_2) 
\end{bmatrix}.$ 
\end{center}
Therefore, $\vec{y}$(0) is the vector that contains the $u_i$ (for $i$=1,...,5) percent change of the initial population sizes of the compartments. \\
\indent Because we want to minimize the population of infected people ($I$), we define the objective function $I(\vec{x}, \vec{p}) = \int^T_0 g(\vec{x}, t, \vec{p}) dt = \int^T_0 \dot{I} dt$. We want to analyze this, because we are interested in how much the infected population ($I$) is affected by small changes in the parameters and the initial population size of each compartment ($\vec{p}$). \\
\indent Following the second step of the algorithm for computing the sensitivity equations \cite{BradleyPDE}, the adjoint is $g_x + \lambda^T(F_x-\dot{F}_{\dot{x}})-\dot{\lambda}^T F_{\dot{x}} = 0$ where: 
\begin{itemize}
\item $\lambda$ is the Lagrange multiplier, and $\lambda^T = \begin{bmatrix}
		\lambda_1 & \lambda_2
	\end{bmatrix}$

\item $g_x$ is the partial of $I$ with respect to the population sizes of the compartments ($\vec{x}$)
\begin{center}
$g_x = \frac{\partial\dot{I}}{\partial\vec{x}} = \begin{bmatrix}
	\frac{\partial\dot{I}}{\partial S} & \frac{\partial\dot{I}}{\partial I}
\end{bmatrix} = \begin{bmatrix}
	\alpha I & \alpha S-\mu
\end{bmatrix}$
\end{center}

\item $F_x$ is the partial of our system of ODEs ($F$) with respect to $\vec{x}$
\begin{center}
$F_x = \begin{bmatrix}
	\alpha I + \mu & \alpha S \\
	-\alpha I & -\alpha S + \mu
\end{bmatrix}$
\end{center}

\item $\dot{\vec{x}}$ is the derivative of compartment vector ($\vec{x}$) with respect to time $\left(\dot{\vec{x}}^T = \begin{bmatrix}
	\dot{S} & \dot{I}
\end{bmatrix}\right)$.

\item $F_{\dot{x}}$ is the partial of our system of ODEs ($F$) with respect to the $\dot{\vec{x}}$.
\begin{center}
$F_{\dot{x}} = \begin{bmatrix}
	1 & 0 \\
	0 & 1
\end{bmatrix}$
\end{center}

\item $\dot{F}_{\dot{x}}$ is the parital of the derivative of our ODE system ($\dot{F}$) with respect to $\dot{\vec{x}}$ and $\dot{F}_{\dot{x}} = \begin{bmatrix}
	0 & 0 \\
	0 & 0
\end{bmatrix}$ 
\end{itemize}

\indent Now solving for the $\lambda^T(F_x-\dot{F}_{\dot{x}})$ part of the adjoint, we note that $F_x-\dot{F}_{\dot{x}} = F_x$. Therefore, multliplying the transpose of the Lagrange multiplier vector by the partial of our system of ODEs with respect to the compartmental populations yields:
\begin{center}
$\lambda^T F_x = \begin{bmatrix}
	(\alpha I + \mu)\lambda_1 - \alpha I \lambda_2 & & \alpha S \lambda_1 + (-\alpha S + \mu)\lambda_2
\end{bmatrix}$.
\end{center}
\indent For the final section of the adjoint ($\dot{\lambda}^T F_{\dot{x}}$), the product of the transpose of the Lagrange multiplier vector and the partial of our system of ODEs with respect to the derivative of the compartmental populations is:
$\dot{\lambda}^T F_{\dot{x}} = \begin{bmatrix}
	\dot{\lambda}_1 & \dot{\lambda}_1
\end{bmatrix}$ \\
\indent Combining the results from above we obtain the adjoint equation:
\begin{eqnarray}
\alpha I + (\alpha I + \mu)\lambda_1 - \alpha I \lambda_2 - \dot{\lambda}_1 &=& 0 \nonumber \\
\alpha S - \mu + \alpha S \lambda_1 + (-\alpha S + \mu)\lambda_2 - \dot{\lambda}_2 &=& 0 \nonumber
\end{eqnarray}
where the initial conditions are $\lambda_1 (T) = 0$ and $\lambda_2 (T) = 0$.\\
\indent Now, we simultaneously solve the initial value problem: $F=0$, $\vec{x}(0)^T = \begin{bmatrix} S(0) & I(0) \end{bmatrix}$. In order to find our sensitivity equations, we solve for 
$\frac{dI}{dp}=\int_0^T (g_p+\lambda^TF_p) dt+\lambda^T F_{\dot{x}}|_{t=0} \vec{y}_{x(0)}^{~-1}\vec{y_p}$, where:
\begin{itemize}
\item $g_p$ is the partial of $I$ with respect the the vector of parameters ($\vec{p}$)
\begin{center}
$g_p = \begin{bmatrix}
	0 & IS & -I & 0 & 0
\end{bmatrix}$
\end{center}

\item $F_p$ is the partial of our system of ODEs ($F$) with respect to the the parameters ($\vec{p}$)
\begin{center}
$F_p=\begin{bmatrix}
	-1 & IS & S & 0 & 0\\
	0 & -IS & I & 0 & 0
\end{bmatrix}$
\end{center}

\item $\vec{y}_{p}$ is the partial of the percent change in initial compartment populations ($\vec{y}$) with respect to the parameters ($\vec{p}$)
\begin{center}
$\vec{y}_{p}=\begin{bmatrix} 
	0 & 0 & 0 & -S(0) & 0 \\
	0 & 0 & 0 & 0 & -I(0) 
\end{bmatrix}$
\end{center}

\item $\vec{y}_{x(0)}^{~-1}$ is the inverse of the partial of the percent change in initial compartment populations ($\vec{y}$) with respect to the initial population vector ($\vec{x}$(0))
\begin{center}
$\vec{y}_{x(0)}=\begin{bmatrix} 
	1 - u_1 & 0 \\
	0 & 1-u_2
\end{bmatrix}$,
$\vec{y}_{x(0)}^{~-1}=\begin{bmatrix} 
	\frac{1}{1 - u_1} & 0 \\
	0 & \frac{1}{1-u_2}
\end{bmatrix}$
\end{center}

\item All previously seen expressions ($\lambda^T$, $F_{\dot{x}}$) have the same definitions from above.
\end{itemize}

\indent Within the integrand we have the product of the Lagrange multiplier vector ($\lambda^T$) and the partial of our system of ODEs with respect to the the parameters ($F_p$), which comes to be:
\begin{center}
$\lambda^T F_p = \begin{bmatrix}
	 -\lambda_1 & (\lambda_1-\lambda_2)IS & \lambda_1 S + \lambda_2 I &  0 & 0
\end{bmatrix}$
\end{center}

\indent Multiplying the the Lagrange multiplier vector ($\lambda^T$) with the partial of our system of ODEs with respect to the the derivative of the compartment vector with respect to time ($F_{\dot{x}}$) with the inverse of the partial of the percent change in initial compartment populations with respect to the initial population vector ($\vec{y}_{x(0)}^{~-1}$), and with the partial of the percent change in initial compartment populations with respect to the parameters ($\vec{y}_{p}$), we have: 
\begin{center}
$F_{\dot{x}}|_{t=0} \vec{y}_{x(0)}^{~-1}\vec{y_p}=
\begin{bmatrix}
	0&0&0& \frac{-S(0)}{1-u_1} & 0 \\
	0&0&0& 0 & \frac{-I(0)}{1 - u_2}
\end{bmatrix}$,
\end{center}
\begin{center}
$\lambda^TF_{\dot{x}}|_{t=0} \vec{y}_{x(0)}^{~-1}\vec{y_p} = \begin{bmatrix}
	0&0&0& -\lambda_1\frac{S(0)}{1-u_1} & -\lambda_2\frac{I(0)}{1 - u_2}
\end{bmatrix}$.
\end{center}

\indent Thus, from $\frac{dI}{dp}$ we add $g_p$ and $\lambda^T F_p$ together within the integrand and then add $\lambda^TF_{\dot{x}}|_{t=0} \vec{y}_{x(0)}^{~-1}\vec{y_p}$ outside of it:
\begin{center}
$\frac{dI}{dp} =  \mathop{\mathlarger{\int}}_0^T \begin{bmatrix}
	\begin{bmatrix} 0 & IS & -I & 0 & 0 \end{bmatrix} \\
	+ \begin{bmatrix}  -\lambda_1 & (\lambda_1-\lambda_2)IS & \lambda_1 S + \lambda_2 I &  0 & 0 \end{bmatrix}
\end{bmatrix}dt + \begin{bmatrix} 0 & 0 & 0 & -\lambda_1\frac{S(0)}{1-u_1} & -\lambda_2 \frac{I(0)}{1-u_2}\end{bmatrix}$
\end{center}

\indent Now we have the sensitivity equations:
\begin{eqnarray}
\frac{\partial I}{\partial \Lambda}&=&\int_0^T -\lambda_1dt \nonumber\\
\frac{\partial I}{\partial \alpha}&=&\int_0^T (1+\lambda_1 - \lambda_2) IS dt \nonumber\\
\frac{\partial I}{\partial \mu}&=&\int_0^T \lambda_1 S + (\lambda_2-1) I dt \nonumber\\
\frac{\partial I}{\partial u_1}&=&-\lambda_1\frac{S(0)}{1-u_1} \nonumber\\
\frac{\partial I}{\partial u_2}&=&-\lambda_2\frac{I(0)}{1-u_2} \nonumber
\end{eqnarray}

\subsection*{Comparison of Forward Sensitivity and Adjoint Sensitivity}
Using MATLAB to analyze the sensitivities numerically, we were able to show that both types of sensitivity analyses outlined above produced the same results for this simple system. In our five-dimensional models, it would be necessary to consider many more equations for the forward sensitivity analysis than for the adjoint method. Therefore, we will use the adjoint method for analyzing the sensitivity of both the CVT Model and the SIAD Model.


\section*{Appendix IV}\label{sec:paramest}
\textbf{Estimation of Parameters}\\
This section contains information on how we obtained acceptable ranges for our parameter values.

\subsection*{Estimation of $\Lambda$}
We calculate $\Lambda$, the number of people receiving new prescriptions per month, by taking the new percentage (43\%) of the yearly total of Vicodin prescriptions \cite{volkow2011characteristics, ims2011use} and dividing by the average supply per person, which ranges from 42.7 to 52.8 days \cite{sullivan2008trends}. Then adjusting all of this to fit the time unit of one month, and we arrive at:
\begin{eqnarray}
2,671,212 \leq \Lambda \leq 3,303,044 \nonumber
\end{eqnarray}

\subsection*{Estimation of $\alpha_1$ and $\alpha_2$}
Recall that both $\alpha_1$ and $\alpha_2$ have unit (1/month). In the models, $\alpha_1$ is the rate at which acute medical users of Vicodin become chronic medical users, and $\alpha_2$ is the rate at which acute medical users stop taking Vicodin. The average waiting time for a person in this compartment is $\frac{1}{\alpha_1+\alpha_2}\leq$ 3, based on our definition of the acute medical users being supplied $\leq$ 90 days of Vicodin, the average wait time is less than or equal to 3 months. We use other data and estimations to get a better approximation of the average waiting time. Comparing multiple studies, the average percentage of all opioid patients who have been supplied less than three months is between 63.8\% and 88.7\% \cite{sullivan2008trends}. In the model, the probability of an acute medical user no longer needing Vicodin is $\frac{\alpha_2}{\alpha_1+\alpha_2}$. Therefore, $0.638 \leq \frac{\alpha_2}{\alpha_1+\alpha_2} \leq 0.887$. Through algebraic manipulation, we have:
\begin{eqnarray}
0.638(\alpha_1+\alpha_2) \leq &\alpha_2& \leq 0.339(\alpha_1+\alpha_2). \nonumber 
\end{eqnarray}
Solving for each side of the inequality, we have:
\begin{eqnarray}
0.638\alpha_1+0.638\alpha_2 \leq \alpha_2 &\mbox{and}& \alpha_2 \leq 0.887\alpha_1+0.887\alpha_2 \nonumber \\
0.638\alpha_1 \leq 0.362\alpha_2 &\mbox{and}& 0.113\alpha_2 \leq 0.887\alpha_2 \nonumber \\
1.762\alpha_1 \leq &\alpha_2& \leq 7.850\alpha_1 \nonumber
\end{eqnarray}
\indent Now we need to consider what the average waiting time is in the acute compartment using these ratios. The percentage of acute medical users who have a one-month supply or less of opioids is between 72.9\% and 88.4\% \cite{sullivan2008trends}. Thus, in order to estimate the upper bound of the waiting time, we use the smallest percentage of acute users and the smallest percentage of less than one month supplied. Note that chronic users are passing through $M$, so those patients are in the compartment for the full three months. We assume that the acute users who have more than a one-month supply have a three-month supply.
\begin{eqnarray}
0.638(0.729)(1)+0.638(0.271)(3)+0.362(3) = 2.070 \nonumber
\end{eqnarray}
\indent Therefore, $\frac{1}{\alpha_1+\alpha_2} \leq 2.070$, which yields $0.175 \leq \alpha_1$ when $1.762\alpha_1 \leq \alpha_2$. \\
\indent Next, we get a lower bound on the average waiting time. To do so, we use the largest percentage of acute users and the largest percentage of acute patients with less than one month supplied. We assume those with less than one month supplied received only a one-day supply of Vicodin and those with more than one month supplied received 31 days. The users who become chronic medical users will be in the compartment for 90 days.
\begin{eqnarray}
0.887(0.884)(1)+0.887(0.116)(31)+0.113(90) = 14.14, \nonumber
\end{eqnarray}
and we divide by 30 to get 0.471 months. Thus, $0.471 \leq \frac{1}{\alpha_1+\alpha_2}$. This gives $\alpha_1 \leq 0.240$ when $\alpha_2 \leq 7.850\alpha_1$. Therefore, we have the following system of inequalities:
\begin{eqnarray}
0.175 \leq&\alpha_1&\leq 0.240 \nonumber \\
1.762\alpha_1 \leq&\alpha_2&\leq7.850\alpha_1 \nonumber
\end{eqnarray}

\begin{figure}[h]
\begin{center}
\includegraphics[scale=0.5]{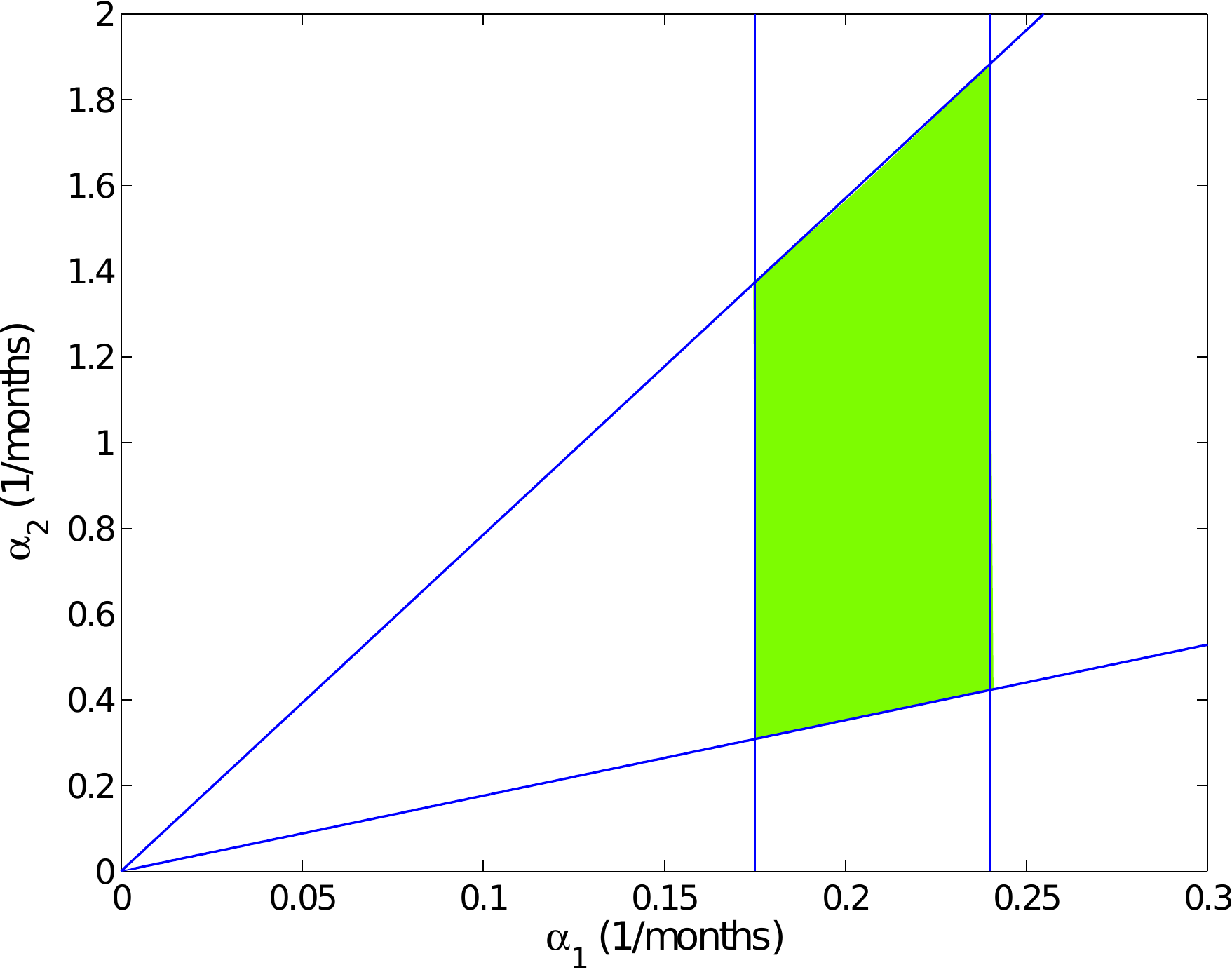}
\caption{\small $\alpha_1$ and $\alpha_2$ Values. \smaller The shaded region indicates the acceptable range of $\alpha_1$ and $\alpha_2$ values.}
\label{AAestimations}
\end{center}
\end{figure}
\indent Realistic parameter estimations for $\alpha_1$ and $\alpha_2$ lie within the shaded area of Figure~\ref{AAestimations}. Starting with the highest intersection point and moving around clockwise, the intersection points are (0.240, 1.884), (0.240, 0.423), (0.175, 0.308), and (0.175, 1.374).

\subsection*{Estimation of $\delta$ and $\beta$}
\indent In the models, $\delta$ (1/month) is the rate at which chronic medical users become abusers, and $\beta$ (1/month) is the rate at which chronic medical users stop using Vicodin. Data suggests that the average opioid exposure time for chronic pain patients ranges between 10.8 and 26.2 months \cite{fishbain2008percentage}. Because these patients will have spent three months as acute patients in the M compartment, the average time in the chronic compartments will be 7.8 to 23.2 months. Therefore, $7.8 \leq \frac{2}{\delta+\beta} \leq 23.2$. From this inequality, we obtain $\delta \leq .256-\beta$ and $\delta \geq .0862-\beta$. \\
\indent Additionally, the percentage of chronic opioid patients who develop abuse ranges from 2.9\% to 11.5\% \cite{edlund2010risks,fishbain2008percentage}. In the model, the probability of chronic medical users becoming abusers is $\left(\frac{\delta}{\delta+\beta}\right)^2$. Thus, $0.029 \leq \left(\frac{\delta}{\delta+\beta}\right)^2 \leq 0.115$. Through algebraic manipulation:
\begin{eqnarray}
0.029(\delta+\beta)^2 \leq &\delta^2& \leq 0.115(\delta+\beta)^2 \nonumber \\
0.170(\delta+\beta) \leq &\delta& \leq 0.339(\delta+\beta) \nonumber
\end{eqnarray}
Solving each side of the inequality, we have:
\begin{eqnarray}
0.170\delta+0.170\beta \leq \delta &\mbox{and}& \delta \leq 0.339\delta+0.339\beta \nonumber \\
0.170\beta \leq 0.830\delta &\mbox{and}& 0.661\delta \leq 0.339\beta \nonumber \\
0.205\beta \leq &\delta& \leq 0.513\beta \nonumber
\end{eqnarray}
We now obtain a system of four inequalities: 
\begin{eqnarray}
\delta &\leq& .256-\beta \nonumber \\
\delta &\geq& .0862-\beta \nonumber \\
\delta &\geq& 0.205\beta \nonumber \\
\delta &\leq& 0.513\beta \nonumber
\end{eqnarray}

\begin{figure}[h]
\begin{center}
\includegraphics[scale=0.5]{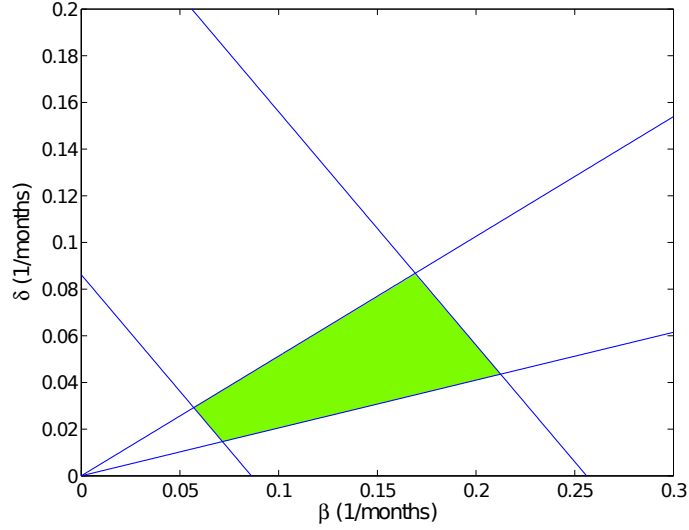}
\caption{\small $\delta$ and $\beta$ Values. \smaller The shaded region indicates the acceptable range of $\delta$ and $\beta$ values.}
\label{BDestimations}
\end{center}
\end{figure}
\indent Realistic parameter estimations for $\delta$ and $\beta$ lie within the shaded area of Figure~\ref{BDestimations}. Starting with the highest intersection point and moving around clockwise, the intersection points are (0.169, 0.0869), (0.213, 0.0436), (0.0715, 0.0147), and (0.0570, 0.0292).

\subsection*{Estimation of $\epsilon$}
We calculate our $\epsilon$ value using the expected value of the time in the abusive compartment, $\frac{1}{\epsilon}$. On average, a person remains an abuser for 24 to 72 months before seeking treatment \cite{mack2003substance}. Using these time frames as bounds for $\frac{1}{\epsilon}$, we determined:
\begin{eqnarray}
0.014 \leq \epsilon \leq 0.042 \nonumber
\end{eqnarray}

\subsection*{Estimation of $\gamma_1$ and $\gamma_2$}
To calculate $\gamma_1$, the rate at which those in the treatment compartment re-enter the abuser compartment, and $\gamma_2$, the successful treatment rate, we defined treatment to last from one to 12 months. We arrive at these bounds from data that relapses occur overwhelmingly within the first year of treatment, in addition to data that indicates treatment should last at least one month \cite{rehabs2013, MHinpatient2013}. We also define relapse to indicate when a person in the treatment compartment returns to pre-treatment abuse levels. For this model, we are not considering an isolated incident of taking a pill to be relapsing, because the amount of time spent in the $A$ compartment and then immediately returning to the $T$ compartment is not relevant. The percentage of people who return to previous abuse levels within one year is 45\%, and the percentage of those who do not return to those levels, which we define as successful treatment, is 55\% \cite{VicodinRehabMH2013}. \\
\indent Because the time in the treatment compartment is defined to be between one and 12 months, we can place upper and lower bounds on the expected waiting time as follows: $1 \leq \frac{1}{\gamma_1 + \gamma_2} \leq 12$. From our known relapse and success percentages, we can express the following: $\gamma_2 = 1.\bar{2}\gamma_1$. Substituting this into the right-hand side of our inequality, we get $\gamma_1 \geq 0.038$. Substituting into the left-hand side, we get $\gamma_1 \leq 0.45$. Thus, the range of values lies along the line $\gamma_2 = 1.\bar{2}\gamma_1$ where $0.038 \leq \gamma_1 \leq 0.450$ \\
\indent For the non-linear models, the units of $\gamma_1$ become 1/(people$\times$month). These units are achieved through dividing the values of $\gamma_1$ by 300 million which is the approximate population of the United States in recent years \cite{census}. Therefore, $1.26\times10^{-10} \leq \gamma_1 \leq 1.50\times10^{-9}$ for the non-linear models. We chose to divide the the original estimate by the total population of the country instead of just the population of Vicodin users considered in the models, because those in treatment can interact with many people that are not considered in the model.

\subsection*{Summary of Estimations}
Note that all parameters have units of (1/month), except $\Lambda$, which has units of (people/month)
\begin{eqnarray}
2,671,212 \leq &\Lambda& \leq 3,303,044 \nonumber \\
0.175 \leq &\alpha_1& \leq 0.240 \nonumber \\
1.762\alpha_1 \leq&\alpha_2&\leq7.850\alpha_1 \nonumber \\
0.014 \leq &\epsilon& \leq 0.042 \nonumber \\
0.046 \leq &\gamma_1& \leq 0.450 \mbox{ \quad\quad\quad~(linear model)} \nonumber \\
1.26\times10^{-10} \leq &\gamma_1& \leq 1.50\times10^{-9} \mbox{ ~~(SIAD model)}\nonumber \\
0.038 \leq &\gamma_2& \leq 0.550 \nonumber
\end{eqnarray}
For $\delta$ and $\beta$ estimations, refer to Figure~\ref{BDestimations}. The parameter values are given in Table~\ref{Parameter Definitions}. 


\subsection*{Excluded Parameters}\label{exparams}
We are able to place a lower bound of 0.00125\%, derived from the number of people who used prescription opioids for non-medical use, and an upper bound of 0.126\%, derived from the number of people who sought treatment for abuse. To obtain the lower bound, we took the number of abusers to be equal to the number of people who have ever used prescription opioids for non-medical users, giving a maximum value for the denominator of our ratio of abusers to overdoses. To obtain the upper bound, we took the number of abusers to be equal to those who sought treatment, minimizing the denominator of the same ratio\cite{CDCPolicyImpact}. We thus concluded that the number of abusers who die from overdosing on Vicodin is not statistically significant and can be neglected for this model. \\
\indent In our model, once a person has exited the population through successful treatment, we do not consider the possibility that the person may re-enter the acute compartment. We make this assumption based on data from requirements of those in treatment \cite{sung2005nonmedical}. \\
\indent When referencing chronic medical users in our model, we are referring specifically to those being treated for conditions that have non-malignant origin. Our research indicates that making this distinction is common, especially when considering prescription drug abuse. These conditions are sometimes referred to as CNCP (chronic non-cancer pain) or NCPC (non-cancer pain condition) \cite{edlund2007risk, edlund2010risks, sullivan2008trends}.\\

\section*{Appendix V}\label{sec:normalizedSensitivityTables}
\begin{table}[htbp]
\caption{\small Percent Change of $A^*$ with respect to $\gamma_1$ (relapse rate)}
\label{Agamma1}
\begin{center}
\begin{tabular}{|c|c|c|c|c|c|c|c|c|c|} \hline
Percent Change of $\gamma_1$ & +1\% & +2\% & +5\% & +10\% & -1\% & -2\% & -5\% & -10\% \\ \hline
Percent Change of A$^*$ & +0.44\% & +0.88\% & +2.22\% & +4.44\% & -0.44\% & -0.88\% & -2.22\% & -4.44\% \\ \hline
\end{tabular}
\end{center}
\end{table}

\begin{table}[htbp]
\caption{\small Percent Change of $A^*$ with respect to $\gamma_2$ (successful treatment rate)}
\label{Agamma2}
\begin{center}
\begin{tabular}{|c|c|c|c|c|c|c|c|c|c|} \hline
Percent Change of $\gamma_2$ & +1\% & +2\% & +5\% & +10\% & -1\% & -2\% & -5\% & -10\% \\ \hline
Percent Change of A$^*$ &  -0.44\% & -0.89\% & -2.22\% & -4.44\% & +0.44\% & +0.88\% & +2.22\% & +4.44\% \\ \hline
\end{tabular}
\end{center}
\end{table}

\begin{table}[htbp]
\caption{\small Percent Change of $A^*$ with respect to $\delta$ (rate of chronic users becoming abusers)}
\label{Adelta}
\begin{center}
\begin{tabular}{|c|c|c|c|c|c|c|c|c|c|} \hline
Percent Change of $\delta$ & +1\% & +2\% & +5\% & +10\% & -1\% & -2\% & -5\% & -10\% \\ \hline
Percent Change of A$^*$ & +1.48\% & +2.96\% & +7.39\% & +14.8\% & -1.48\% & -2.96\% & -7.39\% & -14.8\% \\ \hline
\end{tabular}
\end{center}
\end{table}

\begin{table}[htbp]
\caption{\small Percent Change of $A^*$ with respect to $\beta$ (rate of chronic users ending Vicodin treatment)}
\label{Abeta}
\begin{center}
\begin{tabular}{|c|c|c|c|c|c|c|c|c|c|} \hline
Percent Change of $\beta$ & +1\% & +2\% & +5\% & +10\% & -1\% & -2\% & -5\% & -10\% \\ \hline
Percent Change of A$^*$ & -1.48\% & -2.96\% & -7.39\% & -14.8\% & +1.48\% & +2.96\% & +7.39\% & +14.8\% \\ \hline
\end{tabular}
\end{center}
\end{table}

\begin{table}[htbp]
\caption{\small Percent Change of $A^*$ with respect to $\epsilon$ (rate of abusers entering treatment)}
\label{Aepsilon}
\begin{center}
\begin{tabular}{|c|c|c|c|c|c|c|c|c|c|} \hline
Percent Change of $\epsilon$ & +1\% & +2\% & +5\% & +10\% & -1\% & -2\% & -5\% & -10\% \\ \hline
Percent Change of A$^*$ & -1\% & -2\% & -5\% & -10\% & +1\% & +2\% & +5\% & +10\% \\ \hline
\end{tabular}
\end{center}
\end{table}

\begin{table}[htbp]
\caption{\small Percent Change of $A^*$ with respect to $\Lambda$ (number of newly prescribed Vicodin users per month)}
\label{ALambda}
\begin{center}
\begin{tabular}{|c|c|c|c|c|c|c|c|c|c|} \hline
Percent Change of $\Lambda$ & +1\% & +2\% & +5\% & +10\% & -1\% & -2\% & -5\% & -10\% \\ \hline
Percent Change of A$^*$ & +1\% & +2\% & +5\% & +10\% & -1\% & -2\% & -5\% & -10\% \\ \hline
\end{tabular}
\end{center}
\end{table}


\section*{Appendix VI}
Here we show our derivation of the adjoint equation and sensitivity equation \cite{BradleyPDE}. The sensitivity index component $\frac{\partial A}{\partial p_i}$, for i = 1...$n$ where $n$ is the number of parameters in consideration, is contained in $d_pA$. To compute the total derivative (i.e., gradient) $d_pA = \left(\frac{\partial A}{\partial p_1},\cdot\cdot\cdot\frac{\partial A}{\partial p_{n}}\right) = \int_0^T[g_xd_px+g_p]dt$ we introduce the Lagrangian corresponding to the optimization problem
\begin{eqnarray}
A(\vec{x},\vec{p}) + \int_0^T\lambda^T F(t,\vec{x},\dot{\vec{x}},\vec{p})dt + \mu^T \vec{y}(\vec{x}(0),\vec{p}). \nonumber
\end{eqnarray}
\indent Because $F(t,\vec{x},\dot{\vec{x}},\vec{p}) = 0$, 
\begin{eqnarray}
d_p A = \int_0^T[g_xd_p\vec{x}+g_p]dt + \int_0^T \lambda^T[F_p+F_x d_p \vec{x} + F_{\dot{x}} d_p \dot{x}]dt + \mu^T(\vec{y}_{x(0)} d_p \vec{x}(0) + \vec{y}_p) \nonumber
\end{eqnarray}
where $g_i$, $F_i$, and $\vec{y}_i$ represent partials with respect to $i$ and $\mu^T = \lambda^T F_{\dot{x}} \vec{y}_{x(0)}^{\hspace{.1in}-1}$. \\
\indent Using integration by parts to compute $\int_0^T \lambda^T(F_{\dot{x}} d_p \dot{\vec{x}})dt$ and rearranging terms we get:
\begin{eqnarray}
d_p A &=& \int_0^T(g_p+\lambda^T F_p)dt + \int_0^T[g_x+\lambda^T(F_x-\dot{F}_{\dot{x}}) - \dot{\lambda}^T F_{\dot{x}}]d_p \vec{x} dt \nonumber \\
&&+ \lambda^T F_{\dot{x}}d_p \vec{x} \Big{|}_0^T + \mu^T(\vec{y}_{x(0)}d_p \vec{x}(0) + \vec{y}_p) \nonumber \\
d_p A &=& \int_0^T(g_p+\lambda^T F_p)dt + \int_0^T[g_x+\lambda^T(F_x-\dot{F}_{\dot{x}}) - \dot{\lambda}^T F_{\dot{x}}]d_p \vec{x} dt \nonumber \\
&&+ (\mu^T \vec{y}_{x(0)}-\lambda^T F_{\dot{x}}) \Big{|}_{t=0} d_p \vec{x}(0) + \lambda^T F_{\dot{x}} \Big{|}_{t=T} d_p \vec{x}(T)+\mu^T \vec{y}_p \nonumber
\end{eqnarray}
\indent Due to the fact that $d_p \vec{x}$ is very difficult to calculate, we set $\lambda^T$(T)=0, $\mu^T = \lambda^T F_{\dot{x}} \vec{y}_{x(0)}^{-1}$ and
\begin{eqnarray}
g_x+\lambda^T(F_x-\dot{F}_{\dot{x}}) - \dot{\lambda}^T F_{\dot{x}} = 0 \label{eq:adjoint}
\end{eqnarray}
Now, Equation \ref{eq:adjoint} defines the adjoint equation with $\lambda^T$(T) = 0. Thus:
\begin{eqnarray}
d_p A = \int_0^T (g_p + \lambda^T F_p)dt + \lambda^T F_{\dot{x}} \vec{y}_{x(0)}^{\hspace{.1in}-1} \vec{y}_p \nonumber
\end{eqnarray}
and we just need to solve Equation \ref{eq:adjoint} and compute the appropriate partials to get $d_p A$.


\section*{Appendix VI}
There are many methods of performing sensitivity analysis. However, in order to reduce the number of equations that we are working with, we choose to use the adjoint method. Here we follow the setup of the adjoint method from Bradley \cite{BradleyPDE}. In the end we want to consider the normalized sensitivity indices (i.e $\frac{\partial A}{\partial \delta}\frac{\delta}{A}$) of the abuser population with respect to the parameters in order to see the effects that changing parameter values has on the abuser population ($A$). \\
\indent First, the system of ordinary differential equations for both the CVT (linear) and SIAD (non-linear) Models can be rewritten as vectors equal to zero. For the CVT Model it would be formulated as:
\begin{center}
$\vec{F}(t,$ $\vec{x}$, $\dot\vec{x}$, $\vec{p}) = \begin{bmatrix}
	\dot{M}-\Lambda+(\alpha_1+\alpha_2)M \\
	\dot{C_1}-\alpha_1M+(\delta+\beta)C_1 \\
	\dot{C_2}-\delta C_1+(\delta+\beta)C_2 \\
	\dot{A}-\delta C_2-\gamma_1T+\epsilon A \\
	\dot{T}-\epsilon A+(\gamma_1+\gamma_2) T 
\end{bmatrix} = 0_{5\times1}$.
\end{center}

\indent The initial conditions in both models can be written such that $\vec{y}(\vec{x}(0),\vec{p})= 0$ ($M$(0) - a = 0).
Now, we consider the problem of minimizing $A(\vec{x}, \vec{p})$ where
\begin{eqnarray}
A(\vec{x}, \vec{p}) = \int_0^T g(\vec{x}, \vec{p}, t)dt \nonumber
\end{eqnarray}
subject to $\vec{F}(t,$ $\vec{x}$, $\dot\vec{x}$, $\vec{p}) = 0$ and $\vec{y}(\vec{x}(0),\vec{p})= 0$. \\
\indent  Then we consider the sensitivity index component $\frac{\partial A}{\partial p_i}$, for i = 1...5, is contained in $d_pA$. To compute the total derivative (i.e., gradient) $d_pA = \left(\frac{\partial A}{\partial p_1},\cdot\cdot\cdot\frac{\partial A}{\partial p_14}\right) = \int_0^T[g_xd_px+g_p]dt$ we introduce the Lagrangian corresponding to the optimization problem
\begin{eqnarray}
\mathcal{L} = A(\vec{x},\vec{p}) + \int_0^T\lambda^T F(t,\vec{x},\dot{\vec{x}},\vec{p})dt + \mu^T \vec{y}(\vec{x}(0),\vec{p}). \nonumber
\end{eqnarray}
\indent Because $F(t,\vec{x},\dot{\vec{x}},\vec{p}) = 0$ and $\vec{y}(\vec{x}(0),\vec{p})= 0$ are always satisfied, we are able to set the values of $\lambda$, which depends on time, and $\mu$, which is associated with the initial conditions. Now taking the total derivative, 
\begin{eqnarray}
d_p A = \int_0^T[g_xd_p\vec{x}+g_p]dt + \int_0^T \lambda^T[F_p+F_x d_p \vec{x} + F_{\dot{x}} d_p \dot{x}]dt + \mu^T(\vec{y}_{x(0)} d_p \vec{x}(0) + \vec{y}_p) \nonumber
\end{eqnarray}
where $g_i$, $F_i$, and $\vec{y}_i$ represent partials with respect to $i$. \\
\indent Using integration by parts to compute $\int_0^T \lambda^T(F_{\dot{x}} d_p \dot{\vec{x}})dt$ and rearranging terms, we get:
\begin{eqnarray}
d_p A &=& \int_0^T(g_p+\lambda^T F_p)dt + \int_0^T[g_x+\lambda^T(F_x-\dot{F}_{\dot{x}}) - \dot{\lambda}^T F_{\dot{x}}]d_p \vec{x} dt \nonumber \\
&&+ \lambda^T F_{\dot{x}}d_p \vec{x} \Big{|}_0^T + \mu^T(\vec{y}_{x(0)}d_p \vec{x}(0) + \vec{y}_p) \nonumber \\
d_p A &=& \int_0^T(g_p+\lambda^T F_p)dt + \int_0^T[g_x+\lambda^T(F_x-\dot{F}_{\dot{x}}) - \dot{\lambda}^T F_{\dot{x}}]d_p \vec{x} dt \nonumber \\
&&+ (\mu^T \vec{y}_{x(0)}-\lambda^T F_{\dot{x}}) \Big{|}_{t=0} d_p \vec{x}(0) + \lambda^T F_{\dot{x}} \Big{|}_{t=T} d_p \vec{x}(T)+\mu^T \vec{y}_p \nonumber
\end{eqnarray}
\indent Due to the fact that $d_p \vec{x}$ is very difficult to calculate, we set $\lambda^T$(T)=0, $\mu^T = \lambda^T F_{\dot{x}} \vec{y}_{x(0)}^{-1}$ in order to simplifiy. Also we can avoid computing $d_p \vec{x}$ at all other times $t>0$ by setting
\begin{eqnarray}
g_x+\lambda^T(F_x-\dot{F}_{\dot{x}}) - \dot{\lambda}^T F_{\dot{x}} = 0 \nonumber
\end{eqnarray}
Now, the previous equation defines the adjoint equation with $\lambda^T(T) = 0$. Thus,
\begin{eqnarray}
d_p A = \int_0^T (g_p + \lambda^T F_p)dt + \lambda^T F_{\dot{x}} \vec{y}_{x(0)}^{\hspace{.1in}-1} \vec{y}_p \nonumber
\end{eqnarray}
and we solve the adjoint and compute the appropriate partials to get $d_p A$. This yields the sensitivity equations that, once normalized, we wish to work with.

\listoffigures
\listoftables
\bibliography{GroupBibTeX}
\end{document}